\documentclass[sigconf,screen, nonacm]{acmart}
\AtBeginDocument{%
  }

\usepackage{tabularx}
\usepackage{graphicx}    
\usepackage{subcaption}  
\usepackage{enumitem}
\usepackage{listings}
\usepackage{diagbox}
\usepackage[most]{tcolorbox}
\usepackage{array}
\usepackage{soul}
\usepackage[T1]{fontenc}
\usepackage[utf8]{inputenc}
\usepackage{newtxtext,newtxmath} 
\usepackage{textcomp}
\usepackage{tcolorbox}
\tcbuselibrary{listings, skins,breakable}

\newcolumntype{Y}{>{\raggedright\arraybackslash}X}
\definecolor{promptbg}{RGB}{248,248,248}
\definecolor{promptframe}{RGB}{210,210,210}
\usepackage{float}
\usepackage{xcolor}
\definecolor{ysgreen}{RGB}{46,125,50}      
\definecolor{ysorange}{RGB}{230,140,40} 

\renewcommand\footnotetextcopyrightpermission[1]{}
\setcopyright{none}

\newtcolorbox{promptbox}[1][]{
  enhanced,
  breakable,
  colback=gray!8,
  colframe=gray!50,
  boxrule=0.5pt,
  arc=2pt,
  left=8pt,right=8pt,top=6pt,bottom=6pt,
  fontupper=\small,
  before skip=6pt,
  after skip=6pt,
  #1
}
\lstdefinestyle{jsonstyle}{
  basicstyle=\small\rmfamily,
  breaklines=true,
  breakatwhitespace=false,
  columns=flexible,
  keepspaces=true,
  showstringspaces=false,
  tabsize=2,
  frame=none,
  aboveskip=0pt,
  belowskip=0pt,
}

\newtcblisting{jsonbox}{
  enhanced,
  breakable,
  listing only,
  listing style=jsonstyle,
  colback=gray!8,
  colframe=gray!50,
  boxrule=0.5pt,
  arc=2pt,
  left=8pt,right=8pt,top=6pt,bottom=6pt,
  before skip=6pt,
  after skip=6pt,
}

\author{Yusi Sun}
\authornote{These authors contributed equally to this work.}
\affiliation{%
  \institution{the University of Hong Kong}
  \city{Hong Kong}
  \country{China}
}
\email{soniasun@connect.hku.hk}

\author{Ying Jiang}
\authornotemark[1]
\affiliation{%
  \institution{University of California, Los Angeles}
  \city{Los Angeles}
  \country{United States}
}
\email{anajymua@gmail.com}

\author{Jiayin Lu}
\authornotemark[1]
\affiliation{%
  \institution{University of California, Los Angeles}
  \city{Los Angeles}
  \country{United States}
}
\email{jiayin.kay.lu@gmail.com}

\author{Yin Yang}
\affiliation{%
  \institution{the University of Utah}
  \city{Salt Lake}
  \country{United States}
}
\email{yin.yang@utah.edu}

\author{Yong-Hong Kuo}
\authornote{Corresponding author.}
\affiliation{%
  \institution{the University of Hong Kong}
  \city{Hong Kong}
  \country{China}
}
\email{yhkuo@hku.hk}

\author{Chenfanfu Jiang}
\authornotemark[2]
\affiliation{%
  \institution{University of California, Los Angeles}
  \city{Los Angeles}
  \country{United States}
}
\email{cffjiang@math.ucla.edu}

\begin{document}

\title{\textbf{JARVIS}: A Just-in-Time Augmented Reality VLM-Powered Instruction System for Cross-Reality Task Guidance}

%

\renewcommand{\shortauthors}{Sun et al.}

\begin{abstract}
Many everyday tasks rely on external tutorials such as manuals and videos, requiring users to constantly switch between reading instructions and performing actions, which disrupts workflow and increases cognitive load. Augmented reality (AR) enables in-situ guidance, while recent advances in large language models (LLMs) and vision-language models (VLMs) make it possible to automatically generate such guidance. However, existing AI-powered AR tutorial systems primarily focus on physical procedural tasks and provide limited support for hybrid physical and virtual workspaces. To address this gap, we conduct a formative study of cross-reality tasks and identify key requirements for state awareness and cross-reality coordination. We present \textbf{JARVIS}, a VLM-driven AR instruction system that generates contextual, step-by-step guidance from a single prompt, with real-time state verification and adaptive visual feedback. To inform the system design, we conducted a formative study to understand guidance needs across cross-reality tasks, which we categorize into four types: real-to-real (R2R), real-to-virtual (R2V), virtual-to-real (V2R), and virtual-to-virtual (V2V). A within-subjects study (N=14) across four domains shows \textbf{JARVIS} improves usability, workload, success rate, and visualization effectiveness over baselines. 

\end{abstract}


\begin{teaserfigure}
  \includegraphics[width=\linewidth, trim=0 12bp 0 22bp, clip]{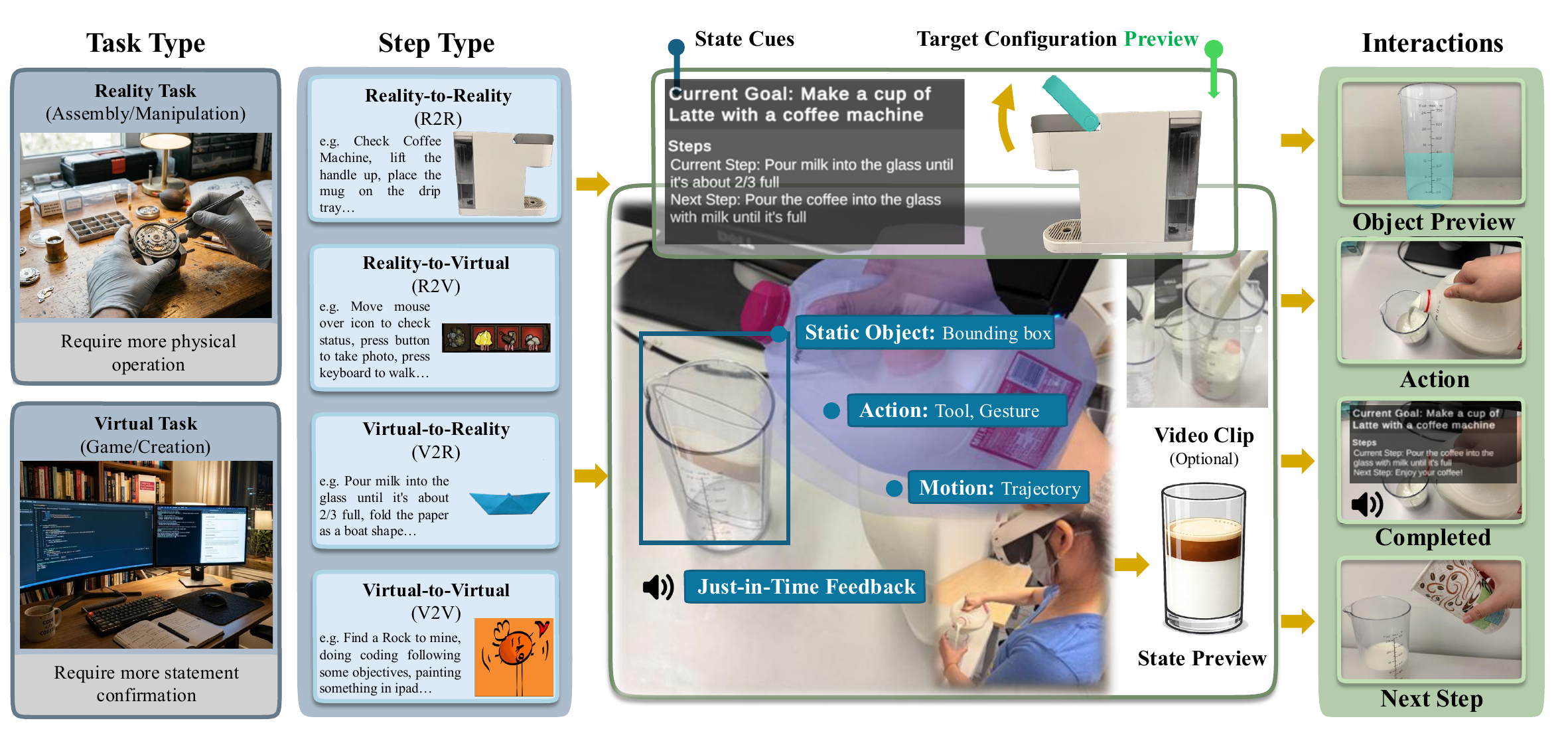}
  \vspace{-25pt}
  \caption{\textbf{JARVIS} instructs a user to make a latte with a coffee machine. Tasks are categorized by step type, Reality-to-Reality (R2R), Reality-to-Virtual (R2V), Virtual-to-Reality (V2R), and Virtual-to-Virtual (V2V), requiring different cross-reality guidance strategies. The system delivers six situated visual guidance types: (1) \textit{State Cues}, showing current goal, active step status, and next step status; (2) \textit{Target Configuration Preview}, overlaying the expected post-step object state, (3) \textit{Static Object Localization} via bounding boxes or masks, (4) \textit{Action Embodiment} rendering segmented tool or hand representations, (5) \textit{Motion Cues} animating trajectory, rotation axis, and direction, and (6) \textit{Just-in-Time Audio Feedback}confirming step completion and advancing to the next step.}
  \vspace{-2pt}
  \Description{The \textbf{JARVIS} system interface shown during a latte-making task. On the left, task types are illustrated with examples: reality tasks such as assembly and manipulation, and virtual tasks such as game operation requiring state confirmation. In the center, the AR view shows a coffee machine scene with an overlaid state panel displaying the current goal, step, and next step, alongside a target configuration preview of a filled glass, a bounding box on the milk container, and an animated trajectory arrow indicating a pouring motion. A speaker icon indicates audio feedback. On the right, interaction outcomes are shown: object preview, action guidance, step completion confirmation, and transition to the next step.}
  \label{fig:teaser}
\end{teaserfigure}

\maketitle
\vspace{-6pt}
\section{Introduction}

Learning real-world tasks, from assembling IKEA furniture to painting, crafting, or configuring digital creative tools, often depends on external instructional materials such as manuals, tutorial videos, and step-by-step demonstrations. However, most instructional materials are labor-intensive to produce, static, and poorly aligned with users’ moment-to-moment needs. They also require frequent switching between instructions and task execution, disrupting workflow and increasing cognitive load. To provide intuitive and seamless instructions, prior work  \cite{mobiletutar22, papertoplace23, processar21} has explored AR devices for in-situ guidance by embedding instructions directly into the user’s workspace, thereby reducing cognitive overhead and supporting situated action. Nevertheless, those AR guidance systems primarily derive contextual guidance from procedural content using either heuristic rules or human demonstrations, requiring task-specific customization. With advances in artificial intelligence (AI), recent work \cite{guided25, xair24} integrates large language models (LLMs) and vision-language models (VLMs) into AR guidance systems to provide more generalizable tutorials by leveraging their knowledge priors.

AI-powered AR guidance systems fall into two paradigms. For one, it provides procedural text-based or symbolic guidance using MLLMs \cite{dogan2024augmented}, LLMs \cite{xair24}, and VLMs \cite{TARGAR24} without requiring pre-registration, while such guidance remains largely text-based and insufficient for complex tasks that require richer multimodal grounding \cite{cooking24}. For another, recent work \cite{guided25, video2coach25} explores multimodal guidance beyond text by using visual cues and state feedback to support spatial understanding and status checks, including visual augmentations such as tool overlays and object highlighting from task manuals \cite{guided25}, as well as progress-aware state feedback and audio guidance from instructional videos \cite{video2coach25}. These approaches focus on physical procedural tasks (e.g., assembly and maintenance), relying on visual cues effective in physical settings, while offering limited support for virtual tasks, where targets are more digital, symbolic, or evolving. Such tasks require more adaptive guidance to convey transformations, intermediate states, and abstract constraints beyond conventional AR tutorials for physical manipulation \cite{gamelearn22, vrgametutorial24}. This motivates us to explore the design of multimodal guidance for diverse tasks across physical and virtual spaces, and to rethink what to visualize, where to present guidance, and how to coordinate workspaces.

To better design multimodal guidance for both virtual and physical tasks, we conducted a formative study comparing three guidance, text, images, and video, across physical (computer setup, origami) and virtual (gameplay, digital painting) tasks. The results show that many tasks are inherently hybrid, spanning both real and virtual domains. Based on this observation, we categorize instructional steps into four types according to the relationship between the guidance source and the locus of action: Real-to-Real (R2R), Real-to-Virtual (R2V), Virtual-to-Real (V2R), and Virtual-to-Virtual (V2V). Moreover, users prefer image guidance for highlighting key steps and video guidance for providing full support in unfamiliar tasks and expect explicit state feedback for progress tracking. These findings inform a design space covering cross-reality steps, state cues, target previews, feedback, action embodiment, motion cues, and object localization.

Based on these findings, we propose \textbf{\textbf{JARVIS}}, an AR visual instruction system that integrates a task planning module and a visual rendering module to generate contextual step-by-step guidance in situ with explicit state feedback from a single prompt. Specifically, given a prompt, a pre-task planner retrieves relevant online materials and produces a structured plan that defines the step sequence, per-step visualization strategies, and localization targets based on the input, current state, and retrieved content. Then an in-task planner performs state verification, sub-planning, and situated rendering. It analyzes the current scene, the active step, and the structured plan to provide context-aware textual and in-situ visual guidance, enabling flexible support when users make errors or misunderstand instructions. Finally, a visual renderer generates multimodal guidance, including images, video clips, motion trajectories, and arrows, along with explicit state feedback. To evaluate the effectiveness and efficiency of the proposed system, we conduct a within-subjects study with 14 participants across R2R, R2V, V2R, and V2V tasks. The results demonstrate that \textbf{\textbf{JARVIS}} provides intuitive and effective guidance, enabling users to complete tasks efficiently across hybrid physical and virtual environments. In summary, our contributions are:

\begin{itemize}
    \item A formative study investigates how AR guidance supports hybrid physical and virtual tasks, and derives a design space for multimodal guidance across cross-reality scenarios.
    
    \item \textbf{\textbf{JARVIS}}, an AR visual instruction system that generates contextual, step-by-step guidance with explicit state feedback to support diverse tutorial tasks from a single prompt.
    
    \item A user study demonstrating the effectiveness and efficiency of the proposed system, as well as its generalizability across a variety of hybrid physical and virtual tasks.
\end{itemize}

\vspace{-6pt}
\section{Related Work}
\vspace{-3pt}
\subsection{Context-Aware Task Guidance}

Research on task guidance has shifted from static, pre-authored instructions to adaptive assistance based on physical situations, action progress, and user intentions \cite{instruar23, adaptuar21, satori25}. Early mixed-reality systems emphasized in-situ and collaborative support for physical tasks. \cite{loki19}. Document-to-AR pipelines \cite{retarget15} showed instruction can be spatially aligned rather than detached, treating context as a core component of interaction rather than a secondary layer. Prior work \cite{processar21, videoar20, mobiletutar22, tutoriallens21} made context operational by explicitly modeling task state, object configuration, and user progress. Video-Annotated AR Assembly Tutorials \cite{videoar20}, ProcessAR \cite{processar21}, TutorialLens \cite{tutoriallens21}, and MobileTutAR \cite{mobiletutar22} connected demonstrations, procedural decomposition, and scene-linked annotations for in-situ step-by-step assistance. However, these systems rely on demonstrations or manual correspondences. Rather than end-to-end text-to-AR transformation, PaperToPlace \cite{papertoplace23} and InstructAR \cite{instructar24} shift attention to contextual grounding: once instruction steps are extracted, the key challenge lies in determining which step is relevant to the current situation, where it should be placed in the scene, and when it should be presented relative to the user’s progress. Beyond adaptation to task state, object configuration, and user progress, some research work \cite{video2coach25, satori25} further explore user- and context-aware adaptation into guidance systems. Vid2Coach \cite{video2coach25} expands guidance to account for user ability, available tools, and alternative execution strategies, while Satori \cite{satori25} moves toward proactive assistance by modeling user belief and environmental context. Despite these advances, existing systems assume instruction and action occur within a single reality layer, limiting cross-reality tasks. Our goal is therefore to support adaptive context-aware guidance across both physical and virtual environments.

\vspace{-6pt}
\subsection{AR Tutorial Systems and Guidance Representations}


AR instruction research shows that effective guidance depends not only on spatial registration, but also on guidance representations \cite{tutoriallens21, mobileaudioanno13}. Early systems such as interactive manuals \cite{asmim18} and authorable AR instructions \cite{arauthorable19} used anchored overlays to deliver procedural content, showing that in-situ presentation reduces attention switching and makes instructions more actionable. However, they rely on manually authored AR elements, limiting support for fine-grained action dynamics. To provide richer procedural information, \cite{tutoriallens21, videoar20, mobileaudioanno13} explored multimodal representations to guide user actions. Audio stickies \cite{mobileaudioanno13}, Video-Annotated AR Assembly Tutorials \cite{videoar20}, TutorialLens \cite{tutoriallens21}, and MobileTutAR \cite{mobiletutar22} explored how demonstrations, captured media, and layered annotations could be organized into usable tutorial flows. This progression suggests that AR tutorials are not simple text-plus-overlay interfaces, but structured combinations of visual elements conveying targets, actions, and temporal order. However, directly presenting demonstrations through images and video might be challenging, due to viewpoint constraints \cite{videoar17}, frequent context switching \cite{papertoplace23}, and limited support for adaptive feedback \cite{adaptuar21, tutoriallens21}. To more effectively leverage multimodal representations, ARticulate \cite{articulate25} explores how motion cues, embodied demonstrations, and calibration-oriented overlays support different forms of procedural understanding, while Guided Reality \cite{guided25} derives a visual guidance taxonomy from user manuals, identifying recurrent forms such as component highlighting, movement indication, hand gesture demonstration, tool-with-motion depiction, and contextual widgets. However, existing guidance designs are largely built around physical procedural tasks, leaving underexplored how visual guidance should be structured when reference, action, or outcome lies in the virtual layer. We therefore propose a cross-reality step taxonomy and a design space to support structured multimodal guidance across hybrid physical and virtual tasks.



\vspace{-6pt}
\subsection{AI for AR Task Guidance}

With advances in artificial intelligence, recent work \cite{xair24, TARGAR24, llmarreview25, xrobject24,caring-ai25,agentar25, guided25, video2coach25, yousri2024illusionx, imaginatear25} increasingly leverage LLMs, VLMs and generative AI to generate procedural steps, infer object references, summarize demonstrations, and provide grounded assistance on demand, reducing reliance on manual authoring. Within this emerging AI-for-AR space, XaiR \cite{xair24} demonstrated that multimodal LLM reasoning can integrate live sensory streams and 3D scene representations for grounded XR assistance, while revealing challenges such as latency and fine-grained spatial alignment. TAGGAR \cite{TARGAR24} addresses part of this gap by using VLMs to derive task guidance from language and images, lacking continuous fine-grained representational planning in situ. XR-Objects \cite{xrobject24} extends this direction to object-centric interaction without pre-registration, while it lacks stepwise guidance. Meanwhile, CARING-AI \cite{caring-ai25} and agentAR \cite{agentar25} address the authoring bottleneck by integrating AI into AR creation workflows, reducing manual efforts. However, they primarily target content creation rather than runtime task support. Guided Reality \cite{guided25} returns to runtime guidance by generating multi-step instructions and spatial anchors, but still relies on staged pipelines for selecting representations and placements. Vid2Coach \cite{video2coach25} further supports execution-time assistance through video-based monitoring and feedback, yet challenges remain in step flexibility. These systems focus on grounding instructions to physical objects and actions, leaving state interpretation, target anticipation, and cross-reality mappings underexplored. In contrast, JARVIS treats AR guidance as both a grounding, a representation, and a planning problem, dynamically determining what to convey, how to present it, and how to adapt to user state and task context.

\vspace{-6pt}
\section{Formative Study}

We present a formative study examining how users perceive and use text, image, and video guidance in both virtual and physical tasks, and use these insights to inform our design space.

\vspace{-10pt}
\subsection{Methods}
\textit{Participants.} {A total of 10 participants (6 males, 4 females) between the ages of 21 and 28 (M=23.9) were recruited. They had varying levels of prior experience across the tasks. Specifically, 2 participants were unfamiliar with the computer task, 7 with the game task, 4 with the origami task, and 3 with the digital painting task.}

\textit{Experiment Design.} The experiment followed a 4 $\times$ 3 (4 Tasks $\times$ 3 Guidance Modalities) within-subjects design to study the effects of guidance modalities on hybrid tasks spanning virtual and physical environments. We considered the following independent variables: (a) \textit{Task.} We included four tasks spanning both physical and virtual domains: computer setup and origami as physical tasks, and gameplay and digital painting as virtual tasks. (b) \textit{Guidance.} We explored three guidance types: text, image, and video. Text provided concise textual step-by-step descriptions. Image guidance presented key steps by highlighting relevant objects with bounding boxes and showing important intermediate states, while video guidance demonstrated full execution with contextual and motion cues. All modalities were designed to be comparable in content coverage across modalities.

\textit{Hypothesis.} We postulated the following three hypotheses: (1). \textbf{H1:} Users’ error rate (\textbf{H1.1}) and completion time (\textbf{H1.2}) will be influenced by the guidance modality of the task. (2). \textbf{H2:} Users will prefer image-based guidance over video, and video over text, as images strike a balance between informativeness and cognitive load, whereas videos provide richer detail with higher cognitive demand, and text offers lower cognitive demand but limited detail.

\textit{Procedures.} We divided the study into two parts. Following prior work, in the first part, participants completed a 100-minute session involving four tasks: computer setup and origami (physical), and gameplay and digital painting (virtual). We prepared text-, image-, and video-based guidance for each task. The text guidance was generated by GPT to resemble real-time textual instructions. The image guidance was created by the research team to approximate visual guidance typically delivered by AR systems. The video guidance consisted of publicly available YouTube videos as well as videos produced by the research team to emulate expert-produced tutorials. For each task, participants were given 20 minutes to complete it under the three guidance modalities, with the order counterbalanced using a Latin square design. Participants were given a 5-minute break after completing each task to mitigate fatigue. All guidance materials were reviewed to ensure that participants could follow them and complete each task. In the second part, participants were asked to complete a questionnaire to collect demographic information, prior task familiarity, and perceived workload using the NASA Task Load Index (NASA-TLX) \cite{nasatlx1988} for each task. They also rated the effectiveness of each guidance modality and their confidence in completing each task without guidance on a 7-point Likert scale. Participants then took part in a 20-minute semi-structured interview to reflect on their experiences with different guidance modalities and task challenges. The interview covered: (a) types of daily tasks requiring guidance; (b) difficulties encountered; (c) modality preferences across real and virtual scenarios; and (d) experiences with state awareness and feedback. 
Combining questionnaire responses with interview data, we conducted a qualitative analysis of participants’ perceptions and expectations of AR guidance across physical and virtual tasks.

\begin{figure}[tbp]
\centering
\includegraphics[width=\linewidth]{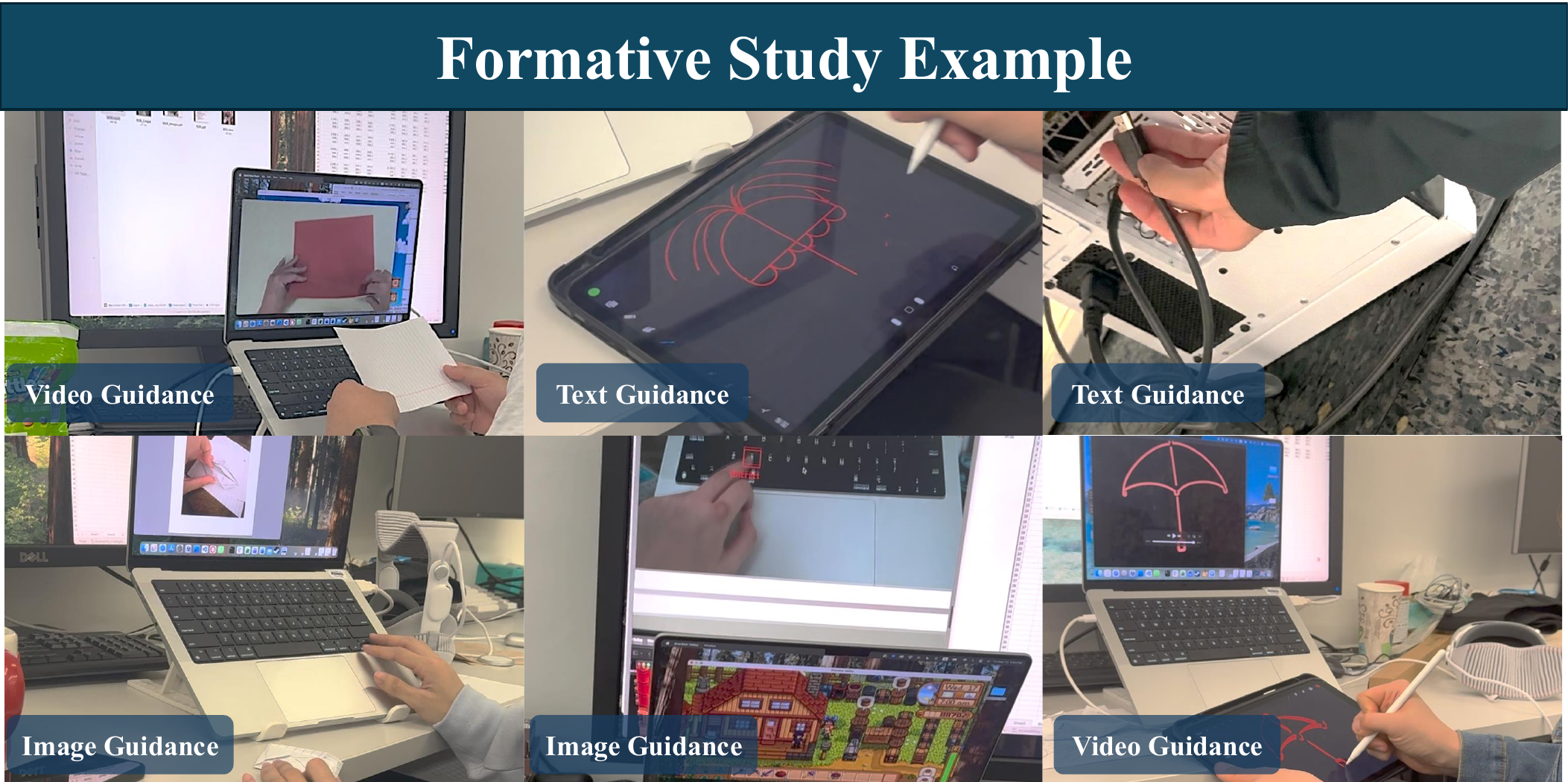}
\vspace{-22pt}
  \caption{Examples from the formative study comparing video, text, and image guidance across representative tasks.}
  \vspace{-10pt}
  \Description{The figure shows six snapshots from the formative study, illustrating how participants followed different guidance modalities—video guidance, text guidance, and image guidance—during hands-on and screen-based tasks.}
  \label{fig:formative}
\end{figure}

\begin{figure}[tbp]
    \centering
    \captionsetup[figure]{skip=5pt}
    \centering
    \includegraphics[width=\linewidth, trim=0 0 0 10bp, clip]{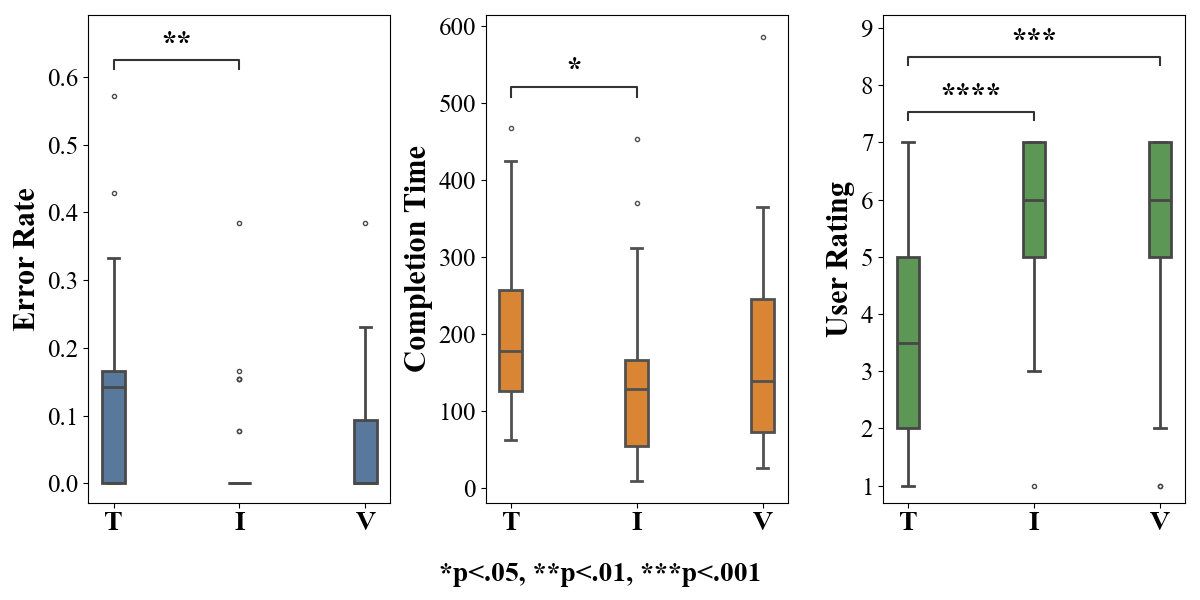}
    \vspace{-22pt}
    \caption{Comparison of error rate, completion time, and user rating across three guidance modalities: Text (T), Image (I), and Video (V). Significant differences are indicated by brackets. }
    \vspace{-10pt}
    \Description{Three side-by-side box plots comparing Text, Image, and Video guidance modalities. The left plot shows error rate, where Text is significantly higher than Image (p < 0.01). The center plot shows completion time, where Text is significantly longer than Image (p < 0.05). The right plot shows user rating, where both Image and Video are rated significantly higher than Text (p < 0.001 and p < 0.0001 respectively). Image and Video guidance consistently outperform Text across all three measures.jyl{Make plot labels larger; Consider making *p etc as labels inside plot}}
    \label{fig:pilotanalysis}
    \vspace{-6pt}
\end{figure}

\vspace{-6pt}            
\subsection{Observation}

All participants completed the tasks. We recorded task execution data and collected completion time, error rate, and step-level errors, along with perceived workload (NASA-TLX), effectiveness, confidence, and task familiarity from questionnaires. Error rate was computed as the proportion of erroneous steps (6, 13, 13, and 7 steps for computer, game, origami, and digital painting, respectively). We used the Friedman test for all measures. Pairwise comparisons were conducted using Wilcoxon signed-rank tests with Holm correction. We report the following observations (O1–O2 on quantitative results and O3–O6 on qualitative insights), which inform our design space. Additional results can be found in the appendix.

\textit{O1 Completion Time and Error Rate.}
Guidance modality significantly affected completion time ($p=0.008$) and error rate ($p < .001$). Text guidance resulted in higher error rates than image guidance (text: M=13\%, SD=1.74 vs. image: M=3\%, SD=1.16, $p=0.001$) and longer completion times (text: M=204.57s, SD=110.84 vs. image: M=130.20s, SD=97.57, $p=0.013$). Although video guidance showed higher error rates (M=5\%, SD=0.55) and longer completion times (M=165.18s, SD=114.08) than image guidance, the differences were not statistically significant. These results indicated that completion time and error rate were influenced by guidance modality, partially supporting \textbf{H1.1} and \textbf{H1.2}. Step-level analysis showed that text guidance introduced ambiguity, while video and image guidance could lead to missed steps due to occlusion or oversight.

\textit{O2 Modality Preference.} 
Fig.\ref{fig:pilotanalysis} showed a significant preference for video (M=5.65, SD=1.34, $p < 0.001$) and image (M=5.425, SD=1.56, $p < 0.001$) over text (M=3.475, SD=1.62), partially supporting \textbf{H2}. The results aligned with user feedback. Seven participants noted that image and video guidance conveyed context more clearly than text.


\textit{O3 Detail Level.} 
Guidance modalities differed in detail. Text conveyed high-level objectives but lacked precision, images balanced key steps and intermediate states, and video provided rich but sometimes redundant detail. This suggested a need for adaptive guidance with concise overviews and on-demand detail.


\textit{O4 Context and Continuity.} 
Video guidance provided continuous context but could be hindered by occlusion, irrelevant motion, or pacing mismatches. All participants struggled to identify which key to press in the gaming task due to occlusion from the fixed camera view in video guidance. Image guidance captured key states without redundancy but lacked transitions for task flow, while text provided minimal context and required users to infer progression. Continuity was treated as a property of the medium rather than the user's current situation, suggesting that real-time state tracking offered a more principled foundation for maintaining contextual continuity.

\textit{O5 Interpretability.} 
Text was often ambiguous, while images better highlighted key objects and actions with more clarity. 
Videos offered richer context but were harder to follow due to occlusion. Participants sought adaptive, environment-aware guidance beyond fixed modalities, with irrelevant content suppressed and concise annotations to improve clarity, capabilities that VLMs could enable.


\textit{O6 Task Characteristics.} Participants reported a diverse set of daily tasks requiring instructional guidance, including 36 real-world tasks and 14 virtual tasks. However, when asked to identify the most difficult tasks, participants selected virtual and real-world tasks in comparable numbers (6 vs. 7), suggesting that virtual tasks, although less frequent, were perceived as equally challenging. Participants showed different preferences for guidance modalities across task types. For real-world tasks, they preferred video-based resources such as social media tutorials and manuals, while for virtual tasks they preferred GPT-like systems that provided more targeted guidance.

\vspace{-6pt}
\subsection{Challenges}
From participants’ behaviors and feedback, we identified key challenges for AR guidance in hybrid physical–virtual tasks.

\textit{Real-to-Virtual and Virtual-to-Real Transitions.} Transitions between real and virtual domains were challenging as users had to verify virtual outcomes of physical actions or translate abstract virtual instructions into precise real-world operations. In real-to-virtual cases, participants struggled to confirm whether a physical action triggered a virtual response, especially with delayed or subtle feedback. For example, P3 described difficulty in a gaming task requiring precise timing and sustained input, where limited feedback made it unclear if the action was correct, while P7 reported uncertainty when plugging in an HDMI cable due to delayed system feedback. In virtual-to-real transitions, participants had to reconcile virtual guidance with their current environment, often without adequate support, as instructions could be abstract, outdated, or insufficiently grounded in the physical context. For instance, P5 followed a navigation guide using an outdated reference image, making it difficult to confirm the correct location, while P11 was unsure how much water to pour in a cooking task due to the lack of a visual reference tied to her specific pot. 



\textit{State Awareness and Feedback.} A common challenge was the lack of clear state awareness and feedback. Participants often reported not knowing which state or mode they were in (P2, P9), especially during cross-reality transitions and virtual tasks where feedback was delayed or absent. P9 noted, “I didn't know what I should do with those icons or tools, and the tutorials online are too long and complex.” Some real-world tasks also required non-visual feedback, such as physical sensations or system responses, for example, sensing car vibrations (P5) or physical discomfort (P11). Given that errors could have serious consequences, participants preferred step-by-step guidance with continuous, reliable feedback.



\textit{Task Complexity and Instruction Mismatch.} Participants reported that difficulties often arose when tutorials lacked sufficient detail or did not match the actual task context. Virtual tasks, though less frequent, were more often perceived as difficult, suggesting higher cognitive demands and variability. Common issues included insufficient instruction detail and mismatches between the tutorial and real situations. These challenges were especially pronounced in cross-reality transitions and fully virtual tasks, where users had to bridge gaps between instruction, execution, and feedback. This pattern highlighted the need for adaptive multimodal guidance and motivated the cross-reality step taxonomy in Section \ref{sec:designspace}.

\begin{figure}[tbp]
    \centering
    \includegraphics[width=\linewidth]{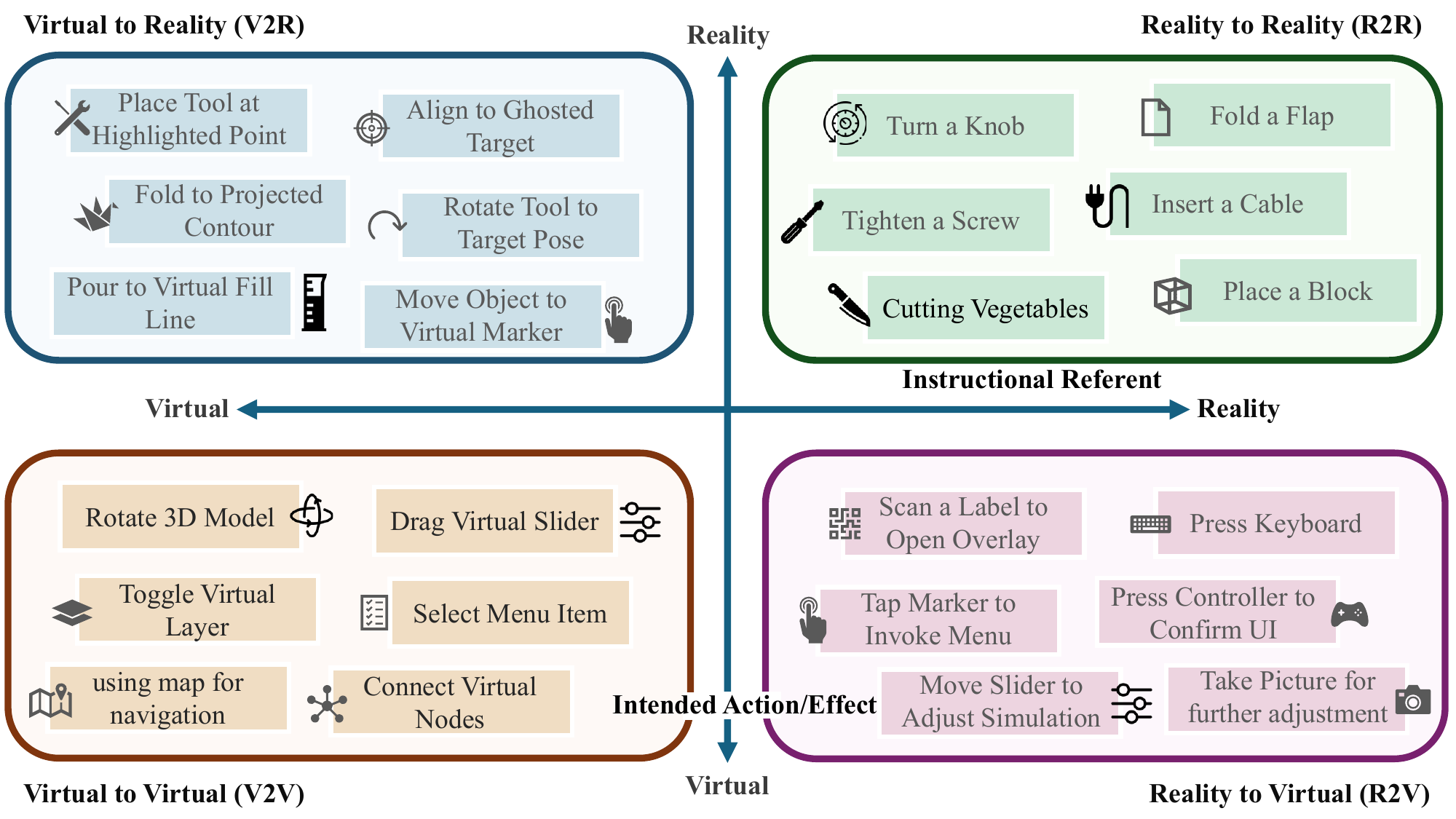}
    \Description{We characterize each instructional step along two dimensions: the location of the instructional referent (real or virtual) and the locus of the intended action or effect (real or virtual). This yields four step types: Real-to-Real (R2R), Real-to-Virtual (R2V), Virtual-to-Real (V2R), and Virtual-to-Virtual (V2V). The figure illustrates how cross-reality tasks can be decomposed into step-level mappings rather than treated as a single task category.}
    \vspace{-17pt}
    \caption{Cross-Reality Step Types. Steps are categorized into four types based on whether the instructional referent and intended action occur in the real or virtual domain.}
\label{fig:steptypes}
\vspace{-15pt}
\end{figure}
\begin{figure}[tbp]
\centering
\begin{subfigure}[t]{0.48\linewidth}
    \centering
    \includegraphics[width=\linewidth]{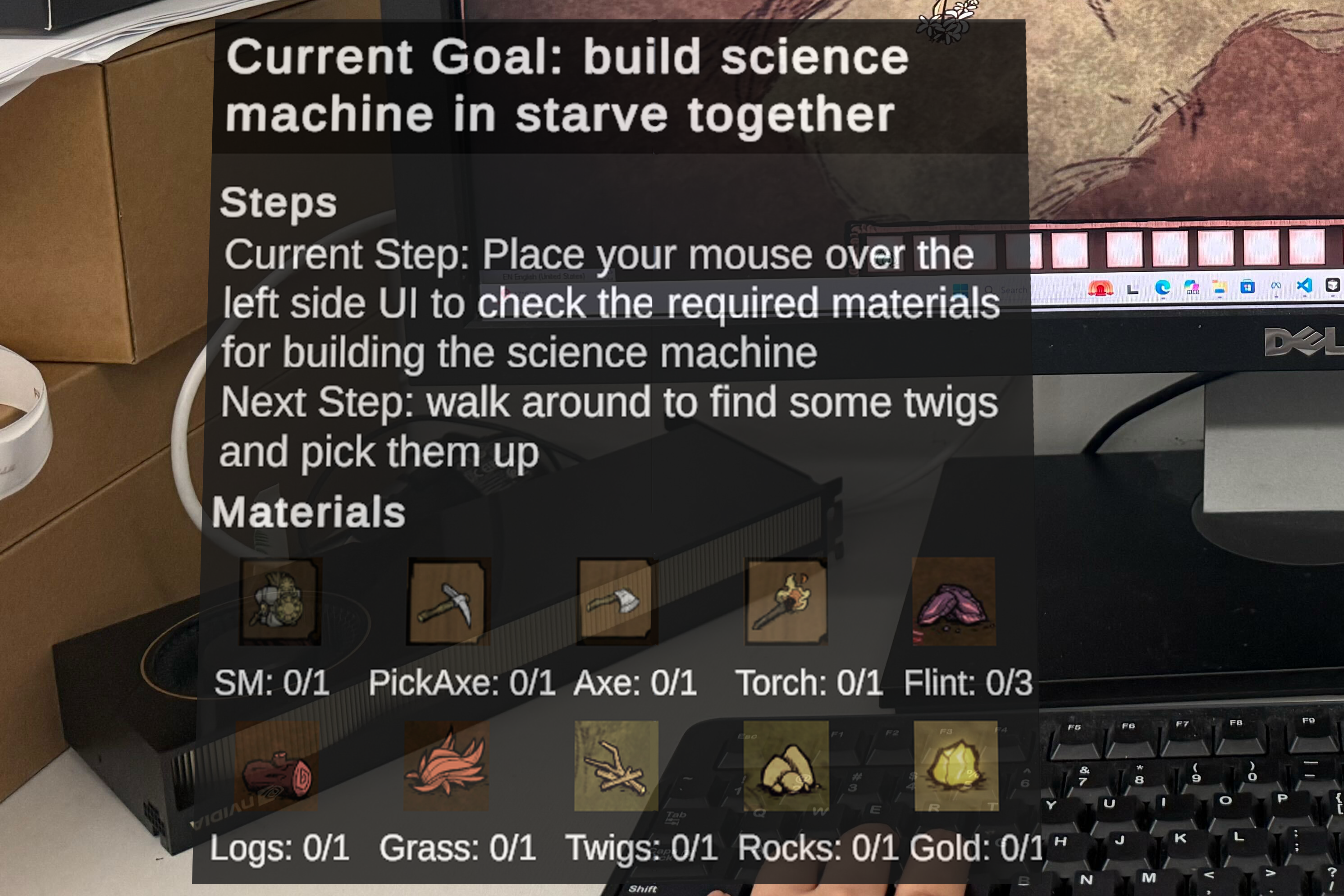}
    \caption{State Cues}
    \label{fig:statepanel}
\end{subfigure}
\hfill
\begin{subfigure}[t]{0.48\linewidth}
    \centering
    \includegraphics[width=\linewidth]{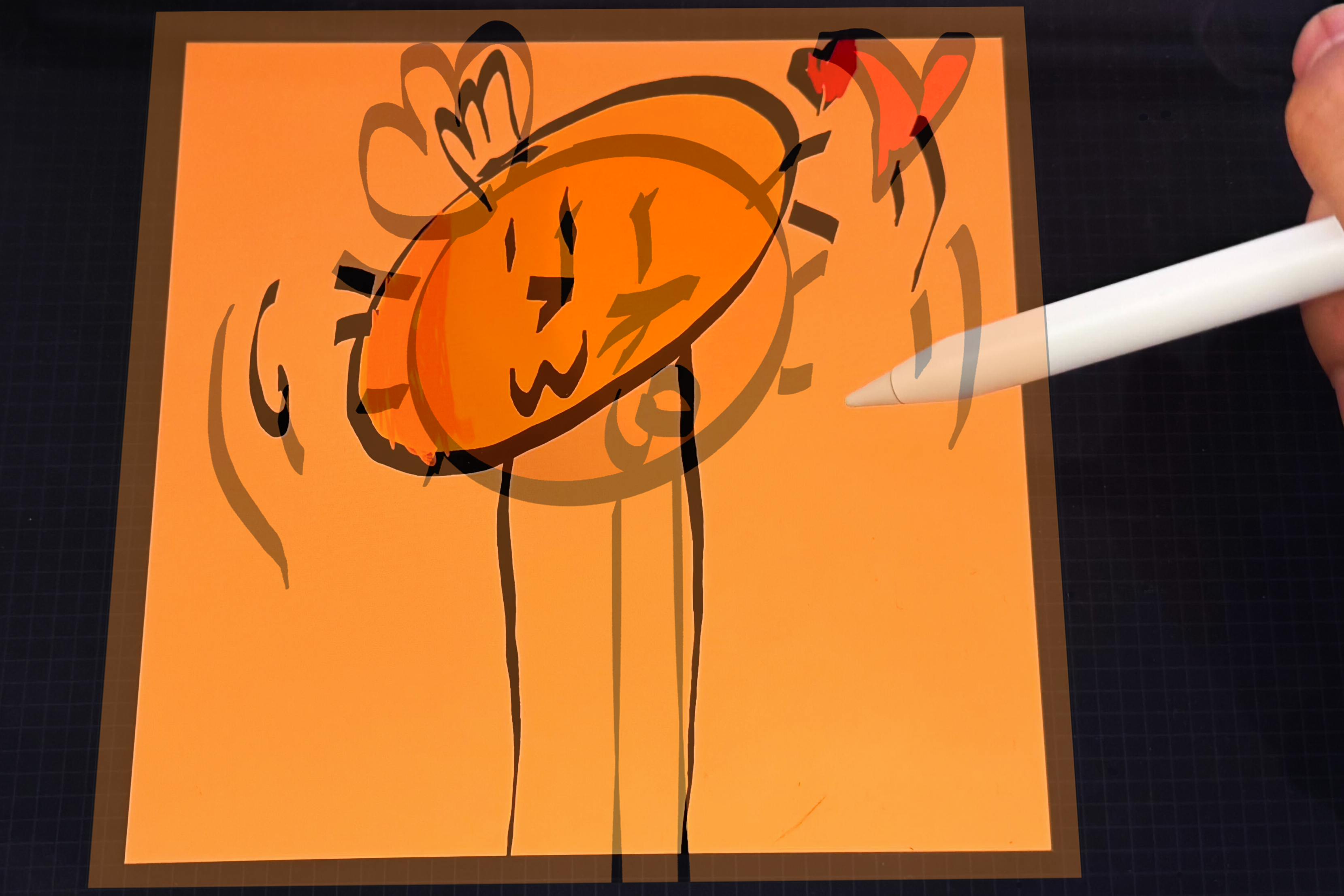}
    \caption{Target Configuration Preview}
    \label{fig:shapePreview}
\end{subfigure}
\vspace{-10pt}
\Description{Two AR guidance components shown side by side. Left (a): a state cues panel overlaid on the scene displaying the current goal, active step instruction, and next step, providing users with a persistent reference for task progress and state transitions. Right (b): a target configuration preview overlaying a semi-transparent silhouette of the expected post-step object configuration onto the physical scene, allowing users to directly compare their current state against the intended outcome.}
\caption{Guidance components: (a) state cues provide feedback on the current state, and (b) target configuration preview visualizes the desired outcome.}
\vspace{-25pt}
\label{fig:state_and_target}
\end{figure}

\vspace{-6pt}
\subsection{Design Space}
\label{sec:designspace}
Based on findings from the formative study, we propose a design space for AR guidance systems that support hybrid tasks.

\textit{D1. Cross-Reality Step Types.} 
Participants reported that many tasks spanned physical and virtual environments with frequent cross-reality transitions, suggesting that treating tasks as purely physical or virtual was insufficient. Thus, we model tasks at the level of individual steps defined by the relationship between the instructional reference and the action as shown in Fig.\ref{fig:steptypes}. Real-to-Real (R2R) refers to steps where both the reference and action occur in the physical world. Real-to-Virtual (R2V) refers to steps where the reference is physical but the effect occurs in the virtual layer. Virtual-to-Real (V2R) refers to steps where the reference is virtual but must be translated into physical actions. Virtual-to-Virtual (V2V) refers to steps where both the reference and action remain in the virtual layer.

\textit{D2. State Cues.} A state panel (Fig.\ref{fig:statepanel}) makes invisible system states explicit during cross-reality task execution. It shows the current step, expected next state, overall goal, and concise subgoals or timers, providing a persistent, at-a-glance reference. This helps resolve ambiguity in R2V and V2V transitions by allowing users to verify whether actions are correctly reflected and in the intended state, consistent with prior findings that explicit state presentation improves situational awareness and task performance \cite{gradual24, arsituaware23}.

\textit{D3. Target Configuration Preview.} A target configuration preview (Fig.\ref{fig:shapePreview}) visualizes the desired end state, enabling users to compare their current configuration with the intended result rather than inferring it from attention cues. This is useful in V2R and V2V steps, where target configurations span position, orientation, geometry, layout, visual state, and parametric values, serving as a feedforward reference for expected outcomes and early mismatch detection, consistent with prior AR guidance research \cite{feedforward24}.

\textit{D4. Correct/Error Feedback.} 
Audio-based correct and error feedback helps users determine whether their current state has reached the next target state, complementing visual cues. This is motivated by findings that visual information alone is often insufficient, especially for non-visible events such as device connections or subtle physical changes, and in cross-reality transitions where physical actions produce hard-to-verify virtual effects. Instruction systems should therefore not only show what to do, but also explicitly confirm correctness, consistent with prior work on corrective feedback \cite{feedforward24}.
\begin{figure*}[tbp]
\centering
\includegraphics[width=\linewidth, trim=0 0 0 10bp, clip]{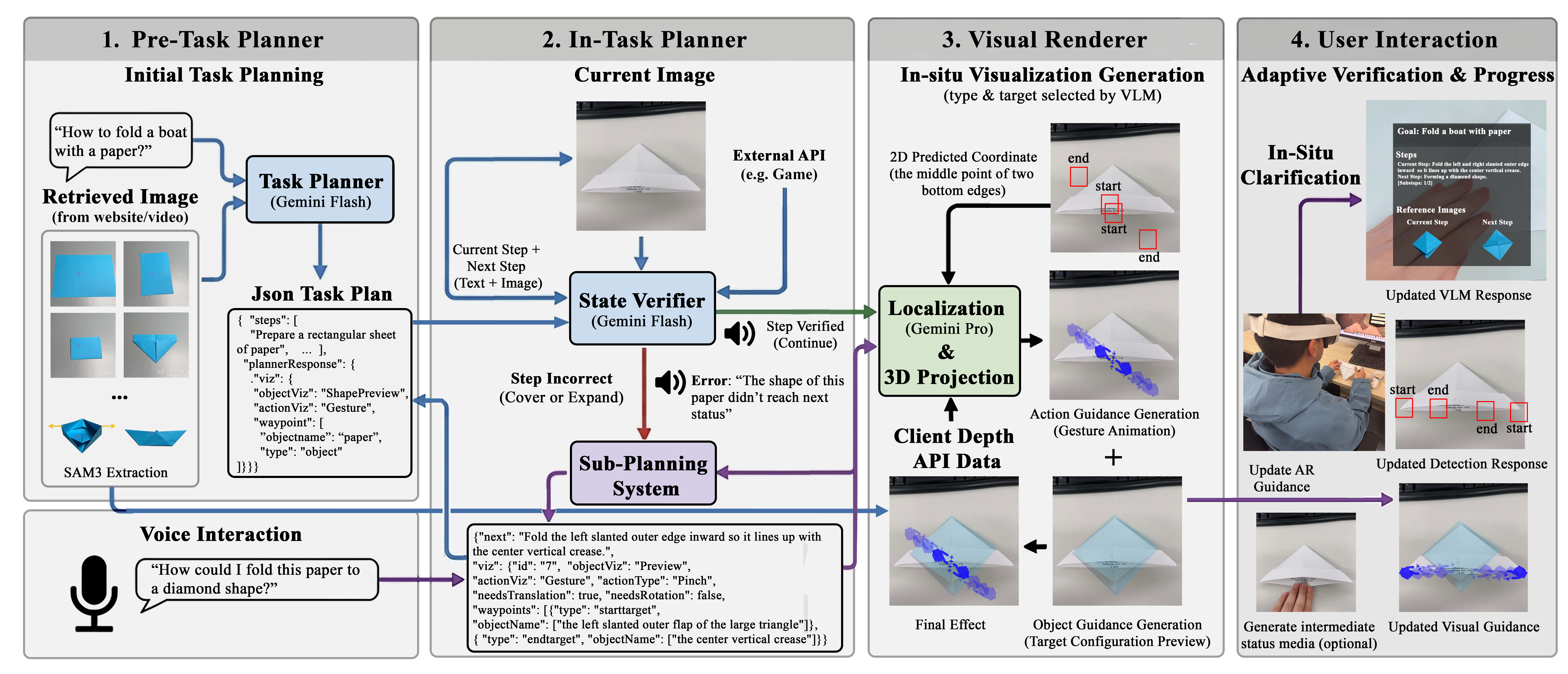}
\Description{A four-panel system architecture diagram. Panel 1 (Pre-Task Planner): a user voice prompt "How to fold a boat with a paper?" feeds into a Gemini Flash Task Planner alongside retrieved tutorial images, SAM3 extracts object silhouettes and the planner outputs a structured JSON plan specifying steps, visualization types, and waypoints. Panel 2 (In-Task Planner): the current scene image and active step are passed to a State Verifier (Gemini Flash), a verified state triggers an audio cue and continues to the Visual Renderer, while an incorrect state, with an error message such as "The shape of this paper didn't reach next status", routes to the Sub-Planning System, which outputs a refined substep JSON. Panel 3 (Visual Renderer): Gemini Pro predicts 2D target coordinates, which are combined with Meta Quest Client Depth API data for 3D projection, the system generates two simultaneous visualization layers, action guidance such as gesture animations with start and end waypoints, and object guidance such as Target Configuration Preview overlays, composited into the final AR effect. Panel 4 (User Interaction): a voice clarification question "How could I fold this paper to a diamond shape?" triggers updated VLM responses, revised object detection, and refreshed AR visual guidance shown on the Meta Quest headset display.}
\vspace{-22pt}
\caption{System architecture of \textbf{JARVIS}, illustrated with an origami boat-folding task, comprising four components: (1) \textit{Pre-Task Planner} generating a structured JSON task plan from user voice input and retrieved images via Gemini Flash and SAM3, (2) \textit{In-Task Planner} verifying task state and either advancing with audio confirmation or invoking sub-planning upon failure, (3) \textit{Visual Renderer} localizing targets via Gemini Pro and projecting guidance into 3D space using Meta Quest depth data, and (4) \textit{User Interaction} supporting in-situ voice clarification with real-time AR updates.}
\label{fig:arch}
\vspace{-16pt}
\end{figure*}

\textit{D5. Action Embodiment.}  
Actions are embodied in AR guidance through hands or tools, illustrating how manipulation should be performed. They may be conveyed through virtual or physical hands to illustrate contact, grasp, or finger placement, or through tool representations spanning physical instruments (e.g., brushes, devices) and virtual or software-based tools (e.g., UI controls, digital brushes, keyboard inputs). This embodiment helps users understand how to perform operations, especially for precise or cross-reality tasks.

\textit{D6. Static Cues.} Static cues indicate target objects (e.g., bounding boxes, masks, highlights), helping users quickly locate relevant regions and support subsequent action understanding.  

\textit{D7. Motion Cues.} 
Unlike static cues, motion cues convey dynamic actions over time (e.g., direction, trajectory, transformation), often using animated arrows or paths. They are especially useful for expressing continuous temporal information, such as changing gestures, tool use, and object state transitions.


\vspace{-10pt}
\section{System Design and Implementation}

In this work, we present \textbf{\textbf{JARVIS}}, a VLM-driven AR instruction system that generates contextual step-by-step guidance in situ with explicit state feedback (Fig.~\ref{fig:arch}). \textbf{JARVIS} treats tutorial execution as a closed-loop process among three components: a \textit{pre-task planner}, which generates a structured step plan and pre-fetches relevant media assets, an \textit{in-task planner}, which continuously verifies the user’s state against expected outcomes and triggers visualization adjustment or sub-step refinement when needed, and a \textit{visual renderer}, which projects step-specific guidance into the physical environment by combining localized object coordinates with VLM-selected visualization strategies. This design enables adaptive step-level refinement and multimodal interaction, allowing guidance to dynamically adjust to user progress and context, improving efficiency and reducing user effort. Our prototype was developed in C\# with Unity and deployed on a Meta Quest 3. A backend server with an NVIDIA RTX 3090 handled VLM inference (Gemini 2.5 Flash and Pro \cite{gemini}) and material retrieval, while a separate server ran SAM3 \cite{carion2025sam} for object localization. The system can run on other machines supporting API-based inference, with sufficient GPU memory required for SAM3.

\vspace{-10pt}
\subsection{Pre-task Planner}

Given a prompt, the pre-task planner generates a structured step plan with visual references and rendering parameters, used by the further in-task planner and visual renderer to deliver continuous, context-aware guidance.



\textit{Initial Planning System.} At session initialization, the user provides an audio prompt, for example, “How to fold a paper boat?”, which the task planner uses to generate a structured high-level plan. To ground each step in visual references, the system retrieves tutorial images and videos via DuckDuckGo \cite{duckduckgo} and yt-dlp \cite{ytdlp}, filters the retrieved content using VLM-based relevance scoring, extracts representative frames from video clips when sequential images are unavailable for certain steps, and applies SAM3 \cite{carion2025sam} to segment target object shapes for AR rendering. The planner then generates a JSON-formatted plan in which each step includes a natural-language instruction, an object localization target, a visualization type selected from the design space, and rendering parameters. Step-level media assets are pre-fetched and associated with each step, enabling efficient reuse during AR rendering without additional latency.


\textit{Voice Interaction.} 
To support in-situ clarification, the system allows users to ask questions via audio at any point during execution. User input is integrated into the current step context, enabling the VLM to generate responses grounded in the active instruction. This allows users to request clarification about unfamiliar components or action details, with responses delivered as updated AR guidance.

\vspace{-10pt}
\subsection{In-task Planner}
After the pre-task planner, the in-task planner performs state verification and sub-planning. It analyzes the current scene, active step, and structured plan to provide context-aware guidance, enabling flexible support when users make mistakes or misunderstand the task.



\textit{State Verification.} The system determines step completion either by receiving an API signal from the software or by querying Gemini-2.5-Flash with the current image, step goal, and verification condition to assess whether the scene matches the expected post-action state. If correct, it advances to the next step and initiates the localization and rendering pipeline for the new instruction. 
If incorrect, it selects a corrective response by revising the visualization strategy (e.g., switching from a bounding box to a motion cue) or invoking sub-planning system to generate a finer-grained recovery sequence.
In addition, correct/error feedback (D4) is provided as an audio cue upon verification, providing an explicit, non-visual confirmation channel when outcomes are not visually observable. This dual-path design addresses two common issues identified in our formative study: lack of state feedback and mismatches between tutorial detail and user needs, mitigated through state-aware updates and adaptive sub-planning.

\textit{Sub-Planning System.} When state verification identifies an incorrect outcome that cannot be resolved by visualization adjustment alone, specifically when the mismatch indicates that the current step requires additional procedural detail, the system invokes the sub-planning component. The sub-planner receives the current step description, the VLM’s failure explanation, and the observed scene image, and decomposes the step into a temporary sequence of intermediate substeps, each with its own instructions, verification conditions, and visualization types. These substeps are inserted into the active state model immediately after the current step, extending the tutorial without disrupting the high-level plan. Once all substeps are verified, the system resumes the original plan. This mechanism is motivated by the observation that many tutorials underspecify difficult steps, and that adaptive local expansion is often more effective than full replanning in most failure cases.

\vspace{-6pt}
\subsection{Visual Renderer}
Based on the verified step specification produced by the in-task planner, we design a visual renderer to provide spatially anchored instruction and state feedback, compositing object state and action overlays directly onto the user’s physical environment in real time.


\textit{Localization.} Once a step is verified as the active target, the system localizes the relevant region using Gemini-2.5-Pro to predict 2D screen-space coordinates of the object or interaction area in the captured image. These coordinates are combined with the saved camera pose and depth data from the Meta Quest environment depth API to project the 2D reference into a stable 3D world position.



\textit{Situated Visual Renderer.} The rendered visualization type depends on the initially planned or revised guidance specification. Each step simultaneously requires two categories of visualization, corresponding to the design space defined in Section \ref{sec:designspace}: \textit{object state visualization}, which communicates what the target object is and what state it should reach, and \textit{action guidance visualization}, which communicates how the user should act upon it. 

\begin{figure*}[tbp]
\centering
\includegraphics[width=\linewidth, trim=0 0 0 5bp, clip]{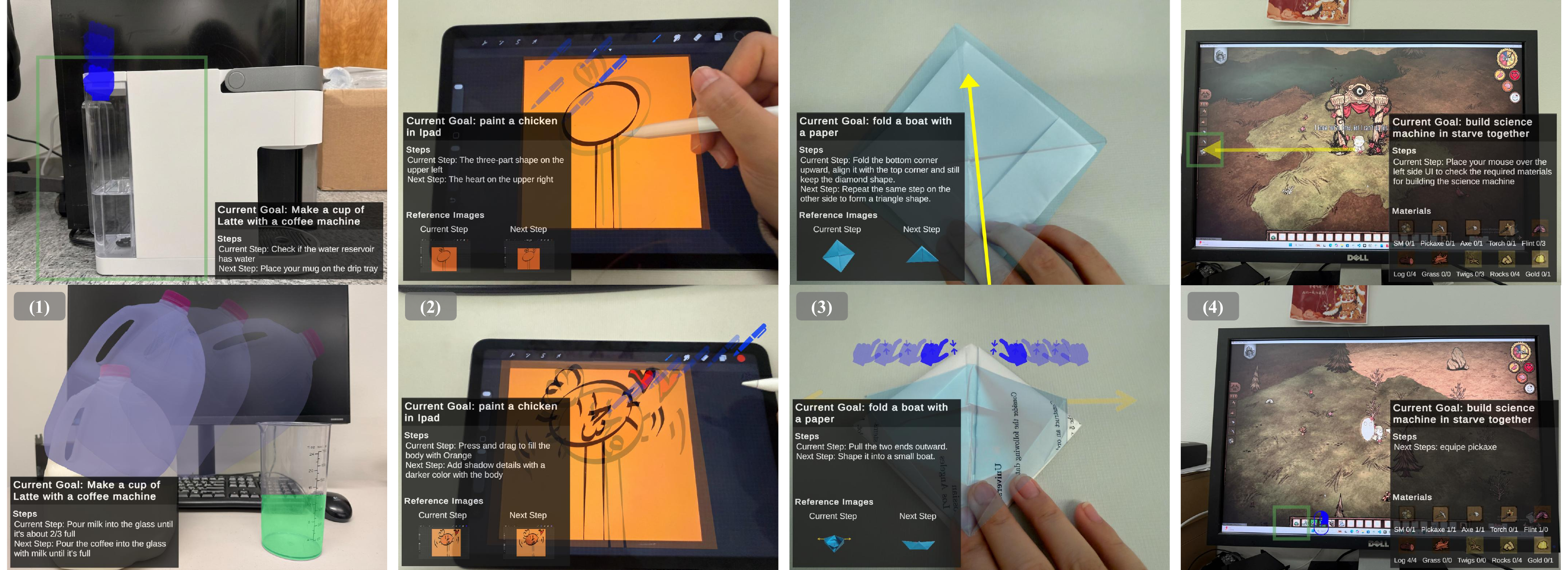}
  \Description{A 3×4 grid of Meta Quest 3 headset screenshots showing \textbf{JARVIS} guidance across four tasks. Coffee task (left column): two views showing a coffee machine scene with a state panel displaying current goal and steps, bounding boxes on the milk container and glass, and a target configuration preview of the filled cup. Digital painting task (center-left column): two views of a tablet drawing interface with gesture overlays indicating brush grip and motion trajectory arrows guiding stroke direction on a digital canvas. Origami task (center-right column): two views of paper folding with a semi-transparent shape preview overlaid on the physical paper and a yellow arrow indicating the fold direction. Gaming task (right column): two views of a game screen with a state cue panel, arrow indicators pointing to relevant UI elements, and reference images showing the target game state.}
  \vspace{-6pt}
  \caption{\textbf{JARVIS} guidance examples across the four user study tasks: (1) coffee machine latte-making with target configuration preview, gesture, motion, tool, and bounding box, (2) digital painting with tool and motion trajectory overlays, (3) origami boat folding with gesture, shape preview and arrow guidance, (4) gaming task with state cues, tool, bounding box and arrow indicators.}
  \label{fig:procedure}
  \vspace{-16pt}
\end{figure*}

\vspace{-6pt}
\section{Technical Evaluation} We conducted a technical evaluation to assess the reliability of \textbf{JARVIS} in query-driven end-to-end pipeline generation and closed-loop recovery. We evaluate whether the system can generate correct step plans, render appropriate visual guidance, verify user state, and recover from failures via revision or sub-planning.


\vspace{-6pt}
\subsection{Methods}
In query-driven evaluation, we collected 8 user queries covering all step types in Section \ref{sec:designspace}, such as “How to change the battery for a controller?”, “How to connect AirPods to a Windows computer?”, and “How to cook an egg?”. For each query, \textbf{JARVIS} generated a structured plan and executed guidance on real or recorded scenes. The set includes 8 tasks and 51 steps. We evaluated the pipeline from planning to rendering, including step correctness, guidance quality, and completion verification. 
As for visual type evaluation, we used previous data to evaluate spatial embedding accuracy under a structured task plan.
We also measure component-wise accuracy and latency for each type.

\vspace{-6pt}
\subsection{Results}
\textit{User Query-Driven Results.} Table \ref{tab:tutorial-quality} reports end-to-end performance on 51 steps from 8 tasks. The system generated correct guidance for 37/51 steps (74.5\%). \textit{Key component} achieved the highest accuracy (90.2\%), while \textit{Image Relevance} was the most challenging category (76.5\%). \textit{Visual Type} selection also remained challenging (80.4\%), indicating limitations in reference retrieval and guidance selection rather than object identification. We observed four main error sources: (1) off-topic retrieval corrupted plans and propagated errors, largely explaining the lower \textit{Image Relevance} scores; (2) limited guidance diversity (e.g., overuse of palm gestures), even when other gesture types or non-gesture guidance would better match the task; (3) highly state-dependent tasks such as origami were particularly difficult to be detected because object geometry changes after each operation; and (4) errors in \textit{state estimation} sometimes prevented the system from advancing at the correct time.

\begin{table}[t]
\centering
\caption{Technical evaluation results across 51 steps from 8 tasks with user queries.}
\vspace{-10pt}
\label{tab:tutorial-quality}
\small
\begin{tabular}{@{}lcccccccc@{}}
\toprule
\textbf{Metric} & \textbf{Total} & \textbf{Correct Steps} & \textbf{Percentage}\\
\midrule
TextInstruction & 51 & 45 & 88.2\% \\
VisualType & 51 & 41 &  80.4\%\\
Key Component & 51 & 46 & 90.2\% \\
Image Relevance & 51 & 39 & 76.5\% \\
Verification & 51 & 42 & 82.3\%\\
\midrule
Target Config Preview & 13 & 11 & 84.6\%\\
Motion & 24 & 20 & 83.3\% \\
Static Object & 38 & 32 & 84.2\%\\
Action & 33 & 28 & 84.8\%\\
\midrule
Total & 51 & 38 & 74.5\%\\
\bottomrule
\end{tabular}
\vspace{-12pt}
\end{table}

\begin{table}[t]
\centering
\caption{Per-type accuracy (Acc.) and latency (Lat.) for visual guidance and component breakdown. 
F denotes Gemini-2.5-Flash and P denotes Gemini-2.5-Pro.}
\vspace{-10pt}
\label{tab:localization}
\small
\begin{tabular}{@{}lcccc@{}}
\toprule
\textbf{Type / Component} & \textbf{Acc. (F)} & \textbf{Lat. (F)} & \textbf{Acc. (P)} & \textbf{Lat. (P)} \\
\midrule
\textbf{Target Config Preview} & 81.1\% & 2.85 & 63.6\% & 20.83 \\
\quad 2D Box & 81.1\% & 2.85 & 63.6\% & 20.83\\
\midrule
\textbf{Motion} & 71.4\% & 2.94 & 85.7\% & 18.23\\
\quad Translation Info & 66.6\% & 2.81 & 83.3\% & 16.19\\
\quad Rotation Info 100\% & 66.67\% & 3.65 & 100\% & 20.23\\
\midrule
\textbf{Static Object} & 54.1\% & 3.83 & 74.2\% &23.29\\
\quad 2D Box &  54.1\% &  3.83 & 74.2\% & 23.29\\
\midrule
\textbf{Action} & 51.6\% & 4.02 & 68\% & 24.31 \\
\quad Tool & 38\% & 4.65 & 66.6\% &28.72\\
\quad Gesture & 80\% & 2.68 & 70\% & 19.89 \\
\midrule
\textbf{Total} & 60.4\% & 3.60 & 71.4\% &22.73 \\
\bottomrule
\end{tabular}
\vspace{-18pt}
\end{table}

\textit{Per-Type Accuracy and Latency.} Table \ref{tab:localization} isolates the visual grounding module and compares two Gemini variants. Gemini-2.5-Pro achieved higher accuracy than Gemini-2.5-Flash (71.4\% vs.60.4\%) but was much slower (22.73s vs.3.60s), revealing a clear accuracy-latency trade-off: Pro is more reliable for complex grounding, while Flash is faster and sufficient for simpler cases. A major error source for both models was recognizing virtual or stylized interfaces. In software (e.g., Photoshop), dense layouts led to frequent mislocalization of icons and controls, while game UIs further increased difficulty due to stylization. Accordingly, performance was generally higher on real-world objects than virtual elements. Performance also varied by guidance type. Flash outperformed Pro on \textit{Target Config Preview} (81.1\% vs.63.6\%), suggesting lightweight models suffice for simple targets. In contrast, Pro performed better on \textit{Motion} (85.7\% and 71.4\%), \textit{Static Object} (74.2\% vs.54.1\%) and \textit{Action} (68.0\% vs.51.6\%), which requires stronger spatial reasoning. Within \textit{Action}, \textit{Tool} localization was the most difficult, especially for Flash (38.0\%).

\vspace{-6pt}
\section{User Study}
We evaluated \textbf{JARVIS} through a controlled within-subjects study, comparing it with two baselines: an \textit{arrow baseline}, which used a static yellow arrow and procedural text to indicate target locations following the design of GuidedReality \cite{guided25}, and an \textit{image baseline}, which presented reference images beside or over the target area using the same retrieved materials and procedural text as JARVIS. We focused on the following research questions:(1). \textbf{RQ1:} Can \textbf{\textbf{JARVIS}} provide higher usability and lower cognitive load compared to conventional guidance baselines? (2). \textbf{RQ2:} Does \textbf{\textbf{JARVIS}} improve task performance relative to arrow-based and image-based guidance? (3). \textbf{RQ3:} Which visual guidance strategies do users find most effective, and how does their utility vary across task contexts?

\begin{figure}[h]
\centering
\includegraphics[width=\linewidth, trim=0 0 0 10bp, clip]{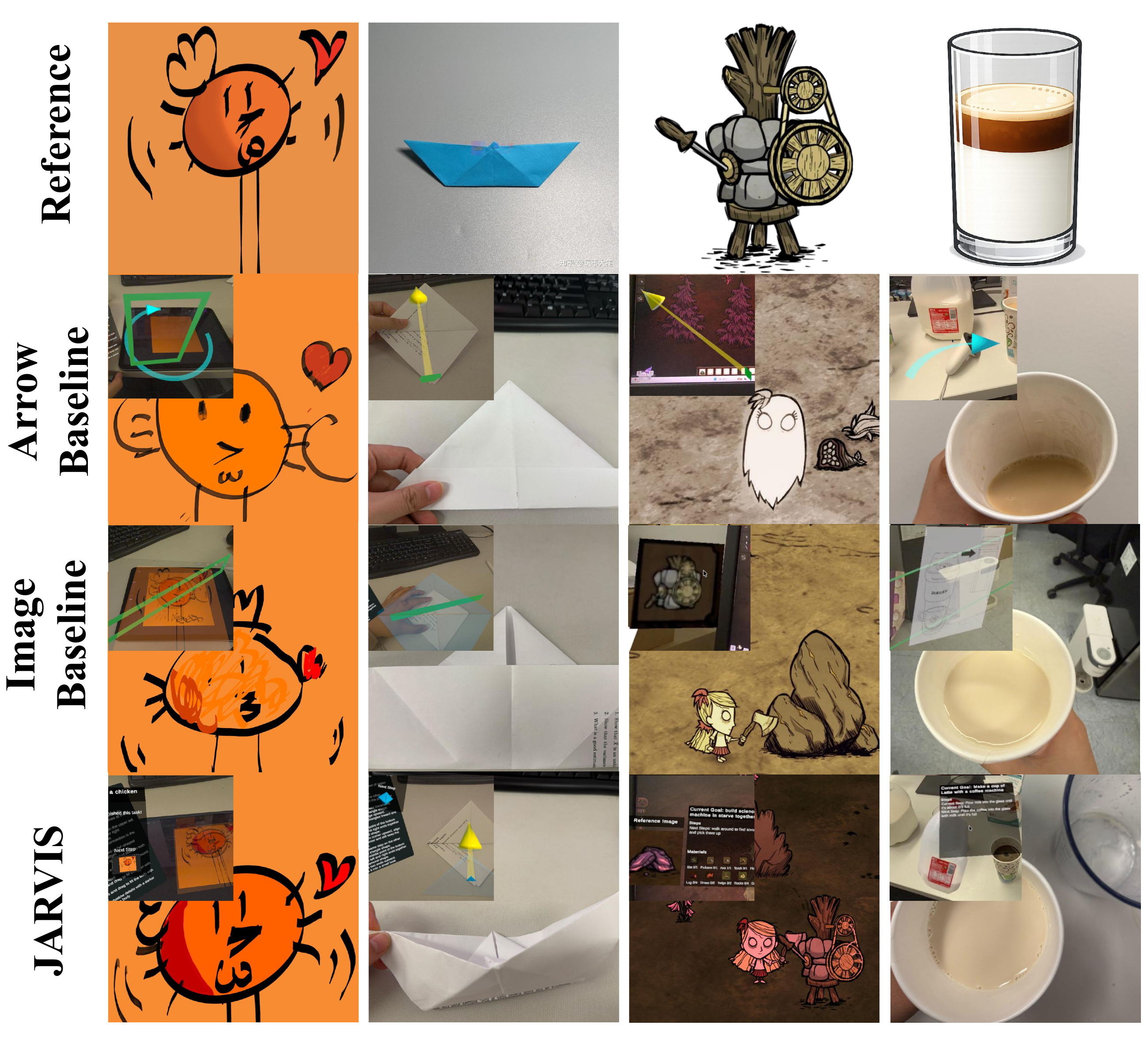}
\vspace{-24pt}
\Description{A 4×4 grid comparing task outcomes. Reference row: target images showing a cartoon character drawing, a paper boat, a game character, and a layered latte. arrow baseline row: users struggled to reproduce the drawing style, produced incorrect folds, failed to navigate the game due to unclear controls leading to character death, and misjudged latte proportions. image baseline row: users had difficulty with color filling techniques, made folding errors, struggled to switch tools at the correct step, and achieved partially correct coffee results. \textbf{\textbf{JARVIS}} row: outcomes most closely resembling the reference across all four tasks, with the Meta Quest 3 AR interface visible in the top-left corner of each cell showing the state panel and visual overlays during execution.}
\caption{Comparison of task outcomes across three guidance conditions. Each column shows the reference alongside results from the \textit{arrow baseline}, image baseline, and \textbf{JARVIS} across four tasks (digital painting, origami, gaming, coffee). The inset shows the corresponding Meta Quest 3 AR view during task execution.}
\label{fig:userstudyexamples}
\vspace{-16pt}
\end{figure}

\vspace{-6pt}
\subsection{Methods}

\textit{Participants.} We recruited 14 participants (4 female, 9 male, 1 non-binary) between the  ages of 21 and 29 (M=24.625).  Participants had varying levels of prior experience across tasks. Specifically, 5 participants were unfamiliar with the coffee task, 11 with the gaming task, 10 with the origami task, and 8 with the digital painting task. 

\textit{Apparatus.} We used a Meta Quest 3 headset as the AR device for the study. The headset was connected to a server with an NVIDIA RTX 3090 GPU to process API calls and fetch real-time results.

\textit{Tasks and Procedure.}
Participants completed six tasks (two per condition) across three conditions: \textbf{\textbf{JARVIS}}, the \textit{arrow baseline}, and the \textit{image baseline}. Following prior work \cite{video2coach25}, we used partial task sampling from four representative hybrid virtual and physical tasks to reduce participant workload: coffee making (R2R), gaming (V2V), origami (V2R), and digital painting (R2V). Task assignment and condition order were counterbalanced to mitigate learning and ordering effects. All participants followed the same pre-generated task plans. Participants were given 10-minute breaks after every three tasks, and sessions were stopped if discomfort was reported. Afterwards, participants filled out a questionnaire on prior task familiarity, system usability (SUS \cite{sus_Brooke_96}), perceived workload (NASA-TLX \cite{nasatlx1988}), and ratings of guidance effectiveness, followed by a semi-structured interview. The entire session lasted around 2 hours. 

\vspace{-6pt}
\subsection{Results}
\begin{figure}[tbp]
    \centering
    \includegraphics[width=\linewidth, trim=0 0 0 26bp, clip]{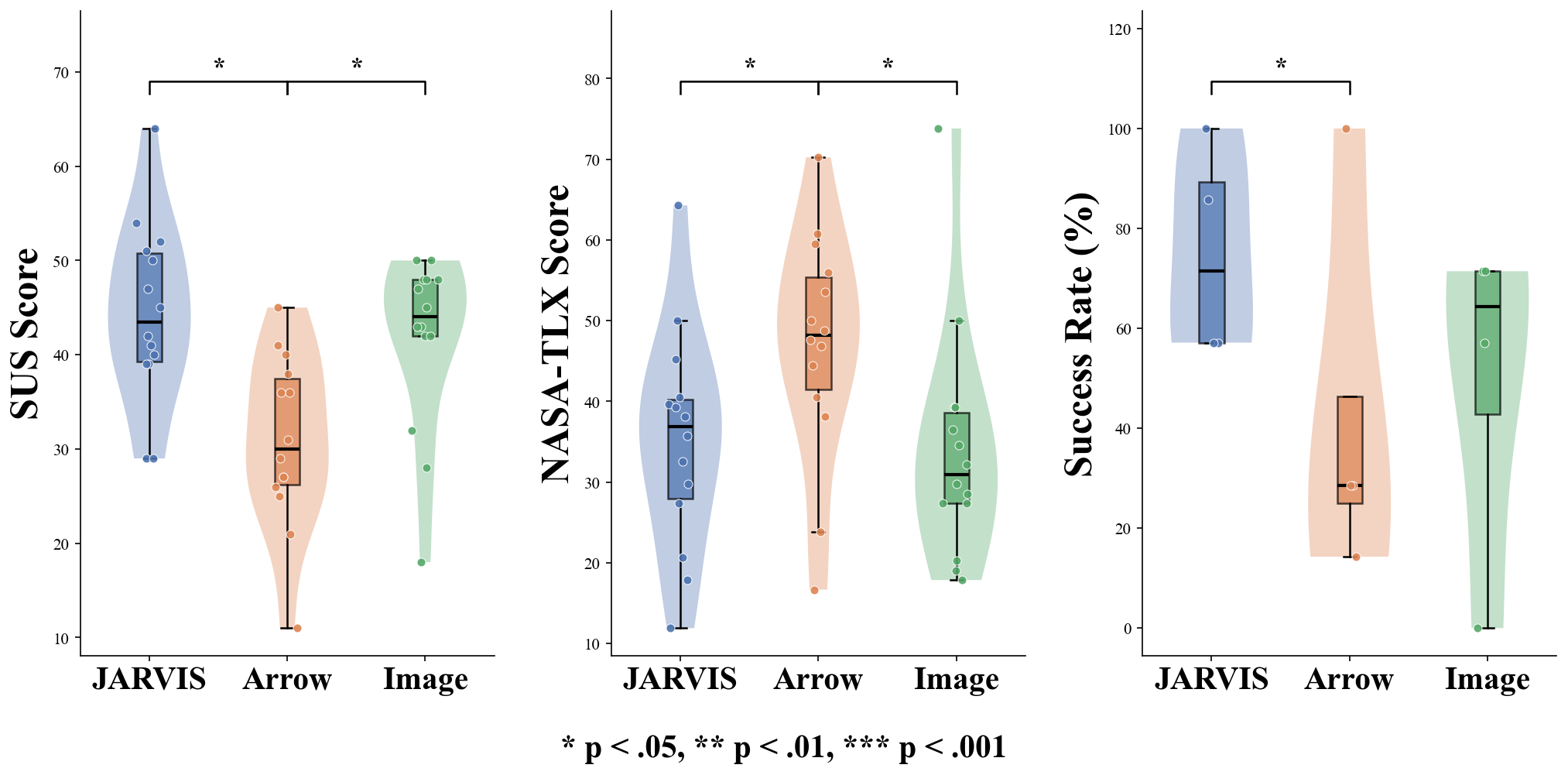}
    \vspace{-22pt}
    \caption{The statistical analysis of different evaluation results. }
    \vspace{-10pt}
    \Description{Error rate, effectiveness, and completion time show significant differences across guidance modalities.}
    \label{fig:staticanalysis}
    \vspace{-10pt}
\end{figure}

We assessed normality using the Shapiro–Wilk test and applied the Friedman test for data analysis. Compared to the two baseline systems, JARVIS demonstrated promising improvements in quantitative measures, including reduced cognitive load as indicated by NASA-TLX scores and improved usability (Fig.\ref{fig:staticanalysis}). Additional metrics and user feedback are presented in the subsections below.



\textit{Usability.}
System condition significantly affected usability. JARVIS  (M=44.43, SD=9.20, p=0.0340) was rated significantly more usable than the \textit{arrow baseline} (M=30.93, SD=8.75, p=0.0226), with no difference from the \textit{image baseline} (M=41.71, SD=9.04, p=0.5289). This aligned with the interview findings. While three participants who preferred the \textit{image baseline} cited greater task familiarity and its more concise layout, 11 out of 14 participants preferred JARVIS. P1 noted that JARVIS helped “bridge gaps between steps,” while P8 found the additional explanations especially useful when reference images did not match the actual scene.



\begin{figure*}[tbp]
    \centering
    \includegraphics[width=.9\linewidth, trim=0 0 0 5bp, clip]{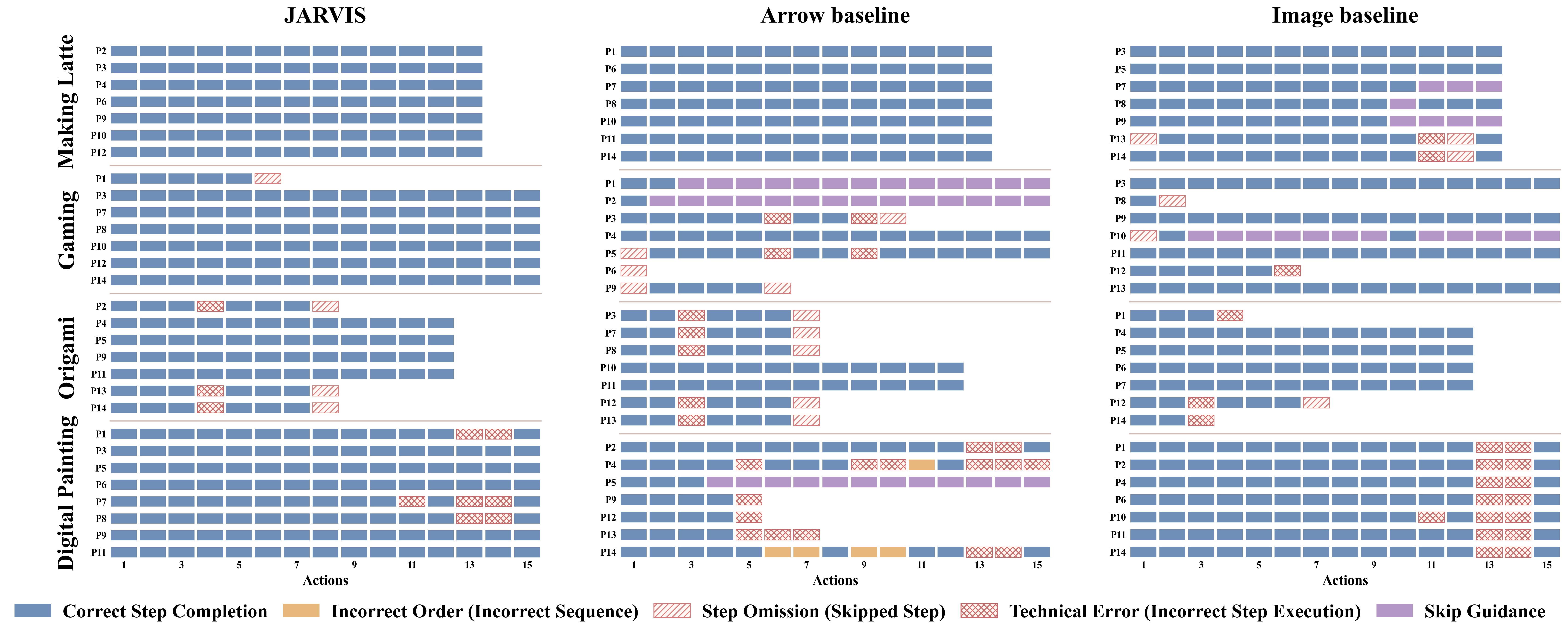}
    \vspace{-5pt}
    \caption{Step-level completion heatmap across three systems (\textbf{\textbf{JARVIS}}, \textit{arrow baseline}, \textit{image baseline}) and four task types (R2R, R2V, V2R, V2V). Each row represents a participant and each column an action step, color-coded by outcome: correct completion, step omission, technical error, incorrect order, and skip guidance.}
    \Description{A 3×4 grid of heatmaps comparing step-level task outcomes across \textbf{\textbf{JARVIS}}, arrow baseline, and image baseline for R2R, R2V, V2R, and V2V task types. \textbf{\textbf{JARVIS}} shows predominantly blue cells across all four task types, with only isolated step omissions in V2R and occasional technical errors in V2V late steps, indicating consistently high completion rates. The arrow baseline shows the most errors overall: R2V and V2V tasks contain dense clusters of skip guidance (purple) and technical errors (crosshatched red), with several participants abandoning mid-task, particularly in R2V where P1, P2, and P6 show extensive non-completion. The image baseline falls between the two, with scattered step omissions and skip guidance events concentrated in R2R late steps and V2V, but fewer technical errors than the \textit{arrow baseline}. Incorrect order errors (orange) appear exclusively in the \textit{arrow baseline} V2V condition. Overall, \textbf{\textbf{JARVIS}} demonstrates the fewest errors and the most consistent step-by-step progression across all task types and participants.}
    \label{fig:finalanalysis}
    \vspace{-6pt}
\end{figure*}

\textit{Perceived Workload.} 
Perceived workload differed significantly across conditions. Both JARVIS (M=35.20, SD=13.07) and the \textit{image baseline} (M=33.99, SD=13.93) yielded lower workload than the \textit{arrow baseline} (M=46.91, SD=13.65, p<0.05), with no difference between JARVIS and the \textit{image baseline} (p=0.7798). These results indicated reduced subjective burden for both JARVIS and the \textit{image baseline} relative to the \textit{arrow baseline}. Overall, these findings partially support \textbf{RQ1}. JARVIS improved perceived usability and reduced workload relative to the \textit{arrow baseline}, but did not significantly outperform the \textit{image baseline}. This indicates that richer, state-aware guidance is more effective than simple directional cues, while offering experience comparable to image-based guidance.


\textit{Task Performance.} The Friedman test found no significant difference in valid completion time across conditions ($\chi^2(2)=1.33$, p=0.5134). The \textit{image baseline} was fastest on average (M=631.33), followed by JARVIS (M=690.42) and the \textit{arrow baseline} (M=707.17)
, likely due to the limited number of participants (n=6) with fully valid paired data.
However, success rate differed significantly: JARVIS achieved the highest success rate (M=0.75, SD=0.21), significantly outperforming the \textit{arrow baseline} (M=0.42, SD=0.34, p=0.0377), while the difference relative to the \textit{image baseline} (M=0.50, SD=0.39) did not reach significance (p=0.0696). Step-level analysis (Fig. \ref{fig:finalanalysis}) reveals a clearer picture. Across all conditions, 88.5\% of steps were completed correctly, while the remaining steps were distributed across technical errors (5.5\%), skip-guidance events (5.9\%, $n=59$), step omissions (2.0\%, $n=20$), and incorrect order errors (0.5\%). JARVIS achieved the best step-level performance, with 96.3\% correct completion (365/379), substantially higher than both the \textit{image baseline} (86.2\%, 294/341) and the \textit{arrow baseline} (74.3\%, 226/304). JARVIS also produced the fewest technical errors (10 cases, 2.6\%) and omissions (4 cases, 1.1\%), and notably had no incorrect-order or skip-guidance events. This suggests that participants were generally able to follow JARVIS continuously and execute the required actions with relatively few breakdowns.
Task-specific patterns further highlight these differences. Procedural tasks like \textit{Making Latte} posed minimal demands and all conditions performed well. Whereas, tasks requiring precise execution (\textit{Origami}), dynamic state tracking (\textit{Gaming}), or sustained multi-step progress (\textit{Digital Painting}) exposed clear baseline limitations: skip-guidance and technical errors accumulated notably, while JARVIS maintained consistently high completion throughout. Overall, JARVIS improves performance by increasing step-level reliability rather than reducing completion time, reducing errors and guidance abandonment for more stable execution (\textbf{RQ2}).

\textit{Effectiveness of Visualizations.}  Perceived effectiveness differed significantly across visualization types ($\chi^2(5)=29.08$, p<0.001). State cues (M=5.4, SD=0.74) and target configuration previews (M=5.33, SD=1.03) were rated highest, as they clearly conveyed task state and completion criteria. Localization-based cues, tool cue (M=3.73, SD=1.23), arrow (M=3.71, SD=1.34), gesture (M=3.61, SD=0.79), and bounding box (M=3.55, SD=1.13), were rated lower, largely due to spatial imprecision where the predicted center drifted away from the target region and small GUI elements further amplified minor localization offsets. Additionally, icon-based tool cues (e.g., left/right-click indicators) were often overlooked or misinterpreted by novices, who treated them as decorative or failed to map them to actions, suggesting that symbolic cues alone were insufficient without state or outcome information. Post-hoc tests showed state cues outperformed the bounding box (p=0.0433) and gesture (p=0.0213), but not the arrow (p=0.0563), tool cue (p=0.0746), or target configuration preview (p=1.000). Target configuration preview also exceeded gesture (p=0.0306). Although differences with the arrow and tool cue were not significant, both were descriptively lower, consistent with reported spatial imprecision. Descriptive statistics revealed task-specific variation in visualization utility. State cues and target configuration previews ranked highest overall, with previews strongest in origami (M=6.12) and digital painting (M=6.38), and state cues leading in coffee (M=5.2) and gaming (M=5.00). In contrast, gesture (gaming, M=2.43), tool cue (origami, M=2.25), and bounding box (digital painting, M=2.88) were least effective, suggesting localization-based cues were sensitive to task precision and cross-reality demands. Overall, participants favored JARVIS’s visualization strategies. Rather than simple object localization, they valued guidance conveying current state, target state, and completion criteria. State cues, target configuration previews, and audio feedback were consistently cited as most effective, highlighting the importance of continuous state-aware and goal-oriented guidance (\textbf{RQ3}).

\textit{Cross-Reality Tutorial System.}
Although SUS scores showed no significant difference between JARVIS and the \textit{image baseline}, 11/14 participants preferred JARVIS. They attributed this to its ability to bridge gaps between steps and provide explanations when reference images did not match the actual scene. For example, P4 stated that it "conveyed the most information clearly and concisely," and P8 highlighted that “the additional explanatory information was especially useful when the images did not match the actual scene.” Several participants also reported difficulty with high-level tasks (e.g., gaming), where instructions were harder to translate into concrete actions. While arrows and highlights helped with localization, they were often too abstract to convey how a step should be performed. Overall, participants favored concise, coarse-grained guidance with optional on-demand detail. 

\begin{figure}[tbp]
\centering
\includegraphics[width=\linewidth, trim=0 0 0 24bp, clip]{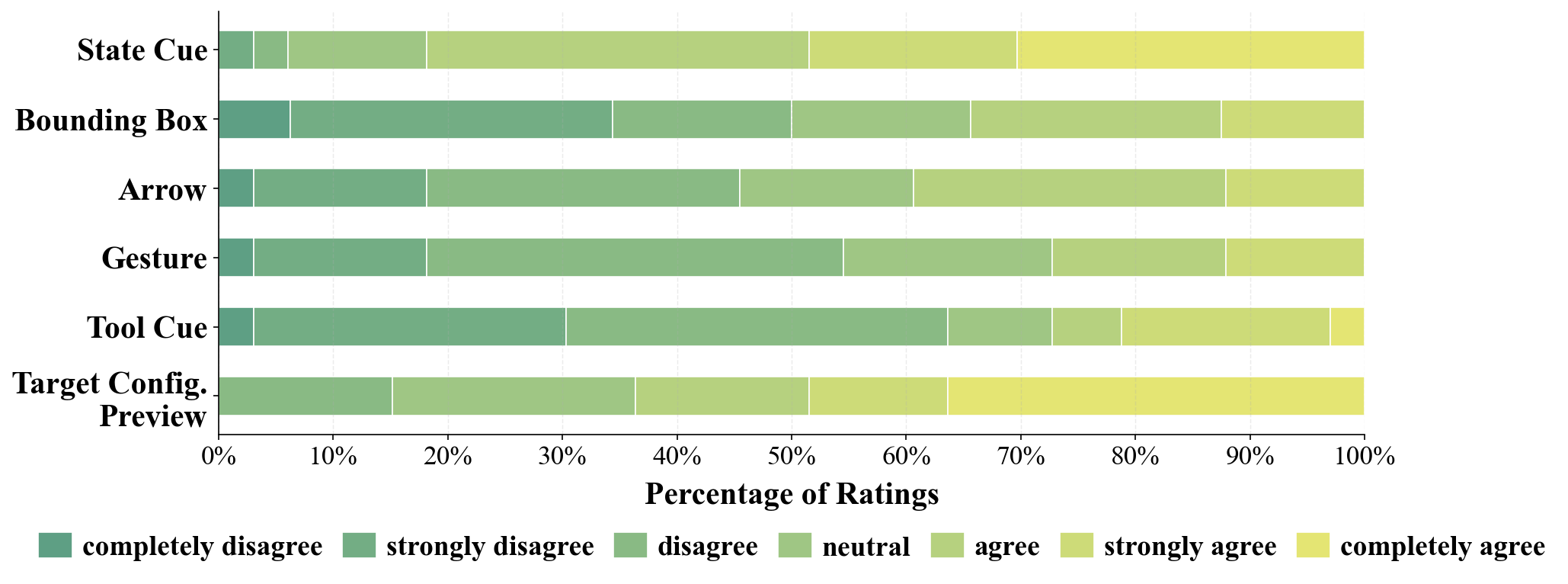}
\vspace{-20pt}
  \caption{Distribution of participants’ ratings on the perceived effectiveness of six visual guidance cues.}
  \Description{Overall, participants reported positive perceptions of all six guidance cues. Gesture, Tool Cue, Arrow, and Bounding Box received predominantly favorable ratings, indicating their effectiveness in supporting action understanding and object localization. State Cue and Target Configuration Preview also showed generally positive evaluations, though with comparatively more neutral responses. These results suggest that participants considered multiple forms of visual guidance helpful for task execution.}
  \label{fig:visualanalysis}
  \vspace{-10pt}
\end{figure}

\vspace{-6pt}
\section{Discussion and Limitations} 
\textit{\textbf{\textbf{JARVIS}} and Baselines in Guiding Users.} 
\textbf{\textbf{JARVIS}} balances informativeness and actionability in cross-reality AR guidance. Quantitative results show that \textbf{\textbf{JARVIS}} achieved the highest usability and success rate, significantly outperforming the \textit{arrow baseline}, while the \textit{image baseline} achieved comparable usability but lower success (\textbf{RQ1, RQ2}). Step-level analysis further shows fewer errors and lower skip rates with \textbf{\textbf{JARVIS}}. While three participants preferred the \textit{image baseline} due to familiarity and lower cognitive load, it could be noticed most users preferred \textbf{\textbf{JARVIS}} for effectively balancing informativeness and cognitive load, valuing clear state, target, and completion cues over the number of visual elements.

\textit{State Interpretation Over Object Localization.} 
As reported, state cues and target previews significantly outperformed localization-based cues, especially in cross-reality tasks. Participants preferred guidance that conveyed state, target configuration, and completion criteria over simple localization. With arrows or bounding boxes alone, users could locate regions but remained uncertain about progress or completion, a limitation amplified in cross-reality steps. Localization cues also suffered from spatial imprecision, particularly for small or dense targets. Their lower performance reflects two key issues: insufficient support for state understanding and limited spatial precision, both addressed by state-oriented visualizations.

\textit{Semantic Gap Between Abstract Cues and Physical Action.} 

While localization cues direct attention, some actions remain hard to convey with icons or arrows alone. Step-level errors (Fig.~\ref{fig:finalanalysis}) illustrate this: press-and-drag actions in digital painting were unclear, right-click cues in gaming were overlooked, and origami folds lacked intermediate states. These results highlight the need for richer representations beyond static cues, like automatically extracted keyframes and explicit object or gesture states from tutorial videos, to better capture action semantics and state transitions.

\textit{Toward Real-Time State Monitoring and Precise Visual Cues.}
Users need more frequent state verification and more precise visual cues. A limitation of \textbf{JARVIS} is state verification is user-triggered rather than continuous, meaning errors may persist until the next interaction. Continuous monitoring could address this but is costly due to many VLM queries. Future work might explore lightweight on-device models for continuous state tracking, reserving VLMs for ambiguous cases to enable real-time error detection with lower overhead. Spatial imprecision also remains a challenge, especially for small GUI elements and fine manipulations. Improved detection or alternative cues (e.g., dynamic animations or contact-point indicators) may enhance precision. Similar issues, such as ambiguous rotation and contact points, have also been reported in Guided Reality\cite{guided25}.

\vspace{-8pt}
\section{Conclusion}

This paper presents \textbf{\textbf{JARVIS}}, an AR instruction system that generates contextual, step-by-step guidance with explicit state feedback to support diverse hybrid virtual and physical tutorial tasks from a single prompt. We conducted a formative study to investigate how AR guidance supports hybrid physical and virtual tasks and to derive a design space for multimodal guidance across cross-reality scenarios. A user study with 14 participants demonstrates that \textbf{\textbf{JARVIS}} achieves higher usability, lower perceived workload, and improved task success rates compared to baselines. Currently, video clips and state feedback are user-triggered. In future work, we aim to incorporate real-time state verification and more precise visual cues to enable timely error detection and more accurate task guidance.



\bibliographystyle{ACM-Reference-Format}
\bibliography{software}


\begin{thebibliography}{39}


\ifx \showCODEN    \undefined \def \showCODEN     #1{\unskip}     \fi
\ifx \showISBNx    \undefined \def \showISBNx     #1{\unskip}     \fi
\ifx \showISBNxiii \undefined \def \showISBNxiii  #1{\unskip}     \fi
\ifx \showISSN     \undefined \def \showISSN      #1{\unskip}     \fi
\ifx \showLCCN     \undefined \def \showLCCN      #1{\unskip}     \fi
\ifx \shownote     \undefined \def \shownote      #1{#1}          \fi
\ifx \showarticletitle \undefined \def \showarticletitle #1{#1}   \fi
\ifx \showURL      \undefined \def \showURL       {\relax}        \fi
\providecommand\bibfield[2]{#2}
\providecommand\bibinfo[2]{#2}
\providecommand\natexlab[1]{#1}
\providecommand\showeprint[2][]{arXiv:#2}

\bibitem[Barquero et~al\mbox{.}(2024)]%
        {cooking24}
\bibfield{author}{\bibinfo{person}{Alexander Barquero}, \bibinfo{person}{Rodrigo~Luis Calvo}, \bibinfo{person}{Daniel~Alexander Delgado}, \bibinfo{person}{Isaac Wang}, \bibinfo{person}{Lisa Anthony}, {and} \bibinfo{person}{Jaime Ruiz}.} \bibinfo{year}{2024}\natexlab{}.
\newblock \showarticletitle{Understanding User Needs for Task Guidance Systems Through the Lens of Cooking}. In \bibinfo{booktitle}{\emph{Proceedings of the 2024 ACM Designing Interactive Systems Conference}} (Copenhagen, Denmark) \emph{(\bibinfo{series}{DIS '24})}. \bibinfo{publisher}{Association for Computing Machinery}, \bibinfo{address}{New York, NY, USA}, \bibinfo{pages}{2006–2018}.
\newblock
\showISBNx{9798400705830}
\href{https://doi.org/10.1145/3643834.3661611}{doi:\nolinkurl{10.1145/3643834.3661611}}


\bibitem[Blattgerste et~al\mbox{.}(2019)]%
        {arauthorable19}
\bibfield{author}{\bibinfo{person}{Jonas Blattgerste}, \bibinfo{person}{Patrick Renner}, {and} \bibinfo{person}{Thies Pfeiffer}.} \bibinfo{year}{2019}\natexlab{}.
\newblock \showarticletitle{Authorable augmented reality instructions for assistance and training in work environments}. In \bibinfo{booktitle}{\emph{Proceedings of the 18th International Conference on Mobile and Ubiquitous Multimedia}} (Pisa, Italy) \emph{(\bibinfo{series}{MUM '19})}. \bibinfo{publisher}{Association for Computing Machinery}, \bibinfo{address}{New York, NY, USA}, Article \bibinfo{articleno}{34}, \bibinfo{numpages}{11}~pages.
\newblock
\showISBNx{9781450376242}
\href{https://doi.org/10.1145/3365610.3365646}{doi:\nolinkurl{10.1145/3365610.3365646}}


\bibitem[Brooke(1996)]%
        {sus_Brooke_96}
\bibfield{author}{\bibinfo{person}{John Brooke}.} \bibinfo{year}{1996}\natexlab{}.
\newblock \showarticletitle{{SUS}: {A} 'Quick' and 'Dirty' Usability Scale}.
\newblock In \bibinfo{booktitle}{\emph{Usability Evaluation in Industry}}, \bibfield{editor}{\bibinfo{person}{Patrick~W. Jordan}, \bibinfo{person}{Bruce Thomas}, \bibinfo{person}{Bernard~A. Weerdmeester}, {and} \bibinfo{person}{Ian~Lyall McClelland}} (Eds.). \bibinfo{publisher}{Taylor and Francis}, Chapter~21, \bibinfo{pages}{189--194}.
\newblock
\showISBNx{9780748404605}


\bibitem[Cao et~al\mbox{.}(2022)]%
        {mobiletutar22}
\bibfield{author}{\bibinfo{person}{Yuanzhi Cao}, \bibinfo{person}{Anna Fuste}, {and} \bibinfo{person}{Valentin Heun}.} \bibinfo{year}{2022}\natexlab{}.
\newblock \showarticletitle{MobileTutAR: a Lightweight Augmented Reality Tutorial System using Spatially Situated Human Segmentation Videos}. In \bibinfo{booktitle}{\emph{Extended Abstracts of the 2022 CHI Conference on Human Factors in Computing Systems}} (New Orleans, LA, USA) \emph{(\bibinfo{series}{CHI EA '22})}. \bibinfo{publisher}{Association for Computing Machinery}, \bibinfo{address}{New York, NY, USA}, Article \bibinfo{articleno}{396}, \bibinfo{numpages}{8}~pages.
\newblock
\showISBNx{9781450391566}
\href{https://doi.org/10.1145/3491101.3519639}{doi:\nolinkurl{10.1145/3491101.3519639}}


\bibitem[Carion et~al\mbox{.}(2025)]%
        {carion2025sam}
\bibfield{author}{\bibinfo{person}{Nicolas Carion}, \bibinfo{person}{Laura Gustafson}, \bibinfo{person}{Yuan-Ting Hu}, \bibinfo{person}{Shoubhik Debnath}, \bibinfo{person}{Ronghang Hu}, \bibinfo{person}{Didac Suris}, \bibinfo{person}{Chaitanya Ryali}, \bibinfo{person}{Kalyan~Vasudev Alwala}, \bibinfo{person}{Haitham Khedr}, \bibinfo{person}{Andrew Huang}, {et~al\mbox{.}}} \bibinfo{year}{2025}\natexlab{}.
\newblock \showarticletitle{Sam 3: Segment anything with concepts}.
\newblock \bibinfo{journal}{\emph{arXiv preprint arXiv:2511.16719}} (\bibinfo{year}{2025}).
\newblock


\bibitem[Chen et~al\mbox{.}(2024)]%
        {vrgametutorial24}
\bibfield{author}{\bibinfo{person}{Boyuan Chen}, \bibinfo{person}{Xinan Yan}, \bibinfo{person}{Xuning Hu}, \bibinfo{person}{Dominic Kao}, {and} \bibinfo{person}{Hai-Ning Liang}.} \bibinfo{year}{2024}\natexlab{}.
\newblock \showarticletitle{Impact of Tutorial Modes with Different Time Flow Rates in Virtual Reality Games}.
\newblock \bibinfo{journal}{\emph{Proc. ACM Comput. Graph. Interact. Tech.}} \bibinfo{volume}{7}, \bibinfo{number}{1}, Article \bibinfo{articleno}{6} (\bibinfo{date}{May} \bibinfo{year}{2024}), \bibinfo{numpages}{19}~pages.
\newblock
\href{https://doi.org/10.1145/3651296}{doi:\nolinkurl{10.1145/3651296}}


\bibitem[Chen et~al\mbox{.}(2023)]%
        {papertoplace23}
\bibfield{author}{\bibinfo{person}{Chen Chen}, \bibinfo{person}{Cuong Nguyen}, \bibinfo{person}{Jane Hoffswell}, \bibinfo{person}{Jennifer Healey}, \bibinfo{person}{Trung Bui}, {and} \bibinfo{person}{Nadir Weibel}.} \bibinfo{year}{2023}\natexlab{}.
\newblock \showarticletitle{PaperToPlace: Transforming Instruction Documents into Spatialized and Context-Aware Mixed Reality Experiences}. In \bibinfo{booktitle}{\emph{Proceedings of the 36th Annual ACM Symposium on User Interface Software and Technology}} (San Francisco, CA, USA) \emph{(\bibinfo{series}{UIST '23})}. \bibinfo{publisher}{Association for Computing Machinery}, \bibinfo{address}{New York, NY, USA}, Article \bibinfo{articleno}{118}, \bibinfo{numpages}{21}~pages.
\newblock
\showISBNx{9798400701320}
\href{https://doi.org/10.1145/3586183.3606832}{doi:\nolinkurl{10.1145/3586183.3606832}}


\bibitem[Chidambaram et~al\mbox{.}(2021)]%
        {processar21}
\bibfield{author}{\bibinfo{person}{Subramanian Chidambaram}, \bibinfo{person}{Hank Huang}, \bibinfo{person}{Fengming He}, \bibinfo{person}{Xun Qian}, \bibinfo{person}{Ana~M Villanueva}, \bibinfo{person}{Thomas~S Redick}, \bibinfo{person}{Wolfgang Stuerzlinger}, {and} \bibinfo{person}{Karthik Ramani}.} \bibinfo{year}{2021}\natexlab{}.
\newblock \showarticletitle{ProcessAR: An augmented reality-based tool to create in-situ procedural 2D/3D AR Instructions}. In \bibinfo{booktitle}{\emph{Proceedings of the 2021 ACM Designing Interactive Systems Conference}} (Virtual Event, USA) \emph{(\bibinfo{series}{DIS '21})}. \bibinfo{publisher}{Association for Computing Machinery}, \bibinfo{address}{New York, NY, USA}, \bibinfo{pages}{234–249}.
\newblock
\showISBNx{9781450384766}
\href{https://doi.org/10.1145/3461778.3462126}{doi:\nolinkurl{10.1145/3461778.3462126}}


\bibitem[Dogan et~al\mbox{.}(2024a)]%
        {xrobject24}
\bibfield{author}{\bibinfo{person}{Mustafa~Doga Dogan}, \bibinfo{person}{Eric~J Gonzalez}, \bibinfo{person}{Karan Ahuja}, \bibinfo{person}{Ruofei Du}, \bibinfo{person}{Andrea Cola\c{c}o}, \bibinfo{person}{Johnny Lee}, \bibinfo{person}{Mar Gonzalez-Franco}, {and} \bibinfo{person}{David Kim}.} \bibinfo{year}{2024}\natexlab{a}.
\newblock \showarticletitle{Augmented Object Intelligence with XR-Objects}. In \bibinfo{booktitle}{\emph{Proceedings of the 37th Annual ACM Symposium on User Interface Software and Technology}} (Pittsburgh, PA, USA) \emph{(\bibinfo{series}{UIST '24})}. \bibinfo{publisher}{Association for Computing Machinery}, \bibinfo{address}{New York, NY, USA}, Article \bibinfo{articleno}{19}, \bibinfo{numpages}{15}~pages.
\newblock
\showISBNx{9798400706288}
\href{https://doi.org/10.1145/3654777.3676379}{doi:\nolinkurl{10.1145/3654777.3676379}}


\bibitem[Dogan et~al\mbox{.}(2024b)]%
        {dogan2024augmented}
\bibfield{author}{\bibinfo{person}{Mustafa~Doga Dogan}, \bibinfo{person}{Eric~J Gonzalez}, \bibinfo{person}{Karan Ahuja}, \bibinfo{person}{Ruofei Du}, \bibinfo{person}{Andrea Cola{\c{c}}o}, \bibinfo{person}{Johnny Lee}, \bibinfo{person}{Mar Gonzalez-Franco}, {and} \bibinfo{person}{David Kim}.} \bibinfo{year}{2024}\natexlab{b}.
\newblock \showarticletitle{Augmented object intelligence with xr-objects}. In \bibinfo{booktitle}{\emph{Proceedings of the 37th Annual ACM Symposium on User Interface Software and Technology}}. \bibinfo{pages}{1--15}.
\newblock


\bibitem[{DuckDuckGo}(2026)]%
        {duckduckgo}
\bibfield{author}{\bibinfo{person}{{DuckDuckGo}}.} \bibinfo{year}{2026}\natexlab{}.
\newblock \bibinfo{title}{DuckDuckGo}.
\newblock \bibinfo{howpublished}{\url{https://duckduckgo.com/}}.
\newblock
\newblock
\shownote{Search engine. Accessed March 30, 2026}.


\bibitem[{Google}(2026)]%
        {gemini}
\bibfield{author}{\bibinfo{person}{{Google}}.} \bibinfo{year}{2026}\natexlab{}.
\newblock \bibinfo{title}{Gemini API}.
\newblock \bibinfo{howpublished}{\url{https://ai.google.dev/api}}.
\newblock
\newblock
\shownote{Accessed: 2026-03-31}.


\bibitem[Hart and Staveland(1988)]%
        {nasatlx1988}
\bibfield{author}{\bibinfo{person}{Sandra~G Hart} {and} \bibinfo{person}{Lowell~E Staveland}.} \bibinfo{year}{1988}\natexlab{}.
\newblock \showarticletitle{Development of NASA-TLX (Task Load Index): Results of empirical and theoretical research}.
\newblock \bibinfo{journal}{\emph{Human mental workload}} \bibinfo{volume}{1}, \bibinfo{number}{3} (\bibinfo{year}{1988}), \bibinfo{pages}{139--183}.
\newblock


\bibitem[Huang et~al\mbox{.}(2021)]%
        {adaptuar21}
\bibfield{author}{\bibinfo{person}{Gaoping Huang}, \bibinfo{person}{Xun Qian}, \bibinfo{person}{Tianyi Wang}, \bibinfo{person}{Fagun Patel}, \bibinfo{person}{Maitreya Sreeram}, \bibinfo{person}{Yuanzhi Cao}, \bibinfo{person}{Karthik Ramani}, {and} \bibinfo{person}{Alexander~J. Quinn}.} \bibinfo{year}{2021}\natexlab{}.
\newblock \showarticletitle{AdapTutAR: An Adaptive Tutoring System for Machine Tasks in Augmented Reality}. In \bibinfo{booktitle}{\emph{Proceedings of the 2021 CHI Conference on Human Factors in Computing Systems}} (Yokohama, Japan) \emph{(\bibinfo{series}{CHI '21})}. \bibinfo{publisher}{Association for Computing Machinery}, \bibinfo{address}{New York, NY, USA}, Article \bibinfo{articleno}{417}, \bibinfo{numpages}{15}~pages.
\newblock
\showISBNx{9781450380966}
\href{https://doi.org/10.1145/3411764.3445283}{doi:\nolinkurl{10.1145/3411764.3445283}}


\bibitem[Huh et~al\mbox{.}(2025)]%
        {video2coach25}
\bibfield{author}{\bibinfo{person}{Mina Huh}, \bibinfo{person}{Zihui Xue}, \bibinfo{person}{Ujjaini Das}, \bibinfo{person}{Kumar Ashutosh}, \bibinfo{person}{Kristen Grauman}, {and} \bibinfo{person}{Amy Pavel}.} \bibinfo{year}{2025}\natexlab{}.
\newblock \showarticletitle{Vid2Coach: Transforming How-To Videos into Task Assistants}. In \bibinfo{booktitle}{\emph{Proceedings of the 38th Annual ACM Symposium on User Interface Software and Technology}} \emph{(\bibinfo{series}{UIST '25})}. \bibinfo{publisher}{Association for Computing Machinery}, \bibinfo{address}{New York, NY, USA}, Article \bibinfo{articleno}{46}, \bibinfo{numpages}{24}~pages.
\newblock
\showISBNx{9798400720376}
\href{https://doi.org/10.1145/3746059.3747612}{doi:\nolinkurl{10.1145/3746059.3747612}}


\bibitem[Keelawat and Suzuki(2024)]%
        {instructar24}
\bibfield{author}{\bibinfo{person}{Panayu Keelawat} {and} \bibinfo{person}{Ryo Suzuki}.} \bibinfo{year}{2024}\natexlab{}.
\newblock \showarticletitle{Transforming Procedural Instructions into In-Situ Augmented Reality Guides with InstructAR}. In \bibinfo{booktitle}{\emph{Adjunct Proceedings of the 37th Annual ACM Symposium on User Interface Software and Technology}} (Pittsburgh, PA, USA) \emph{(\bibinfo{series}{UIST Adjunct '24})}. \bibinfo{publisher}{Association for Computing Machinery}, \bibinfo{address}{New York, NY, USA}, Article \bibinfo{articleno}{70}, \bibinfo{numpages}{3}~pages.
\newblock
\showISBNx{9798400707186}
\href{https://doi.org/10.1145/3672539.3686321}{doi:\nolinkurl{10.1145/3672539.3686321}}


\bibitem[Kong et~al\mbox{.}(2021)]%
        {tutoriallens21}
\bibfield{author}{\bibinfo{person}{Junhan Kong}, \bibinfo{person}{Dena Sabha}, \bibinfo{person}{Jeffrey~P Bigham}, \bibinfo{person}{Amy Pavel}, {and} \bibinfo{person}{Anhong Guo}.} \bibinfo{year}{2021}\natexlab{}.
\newblock \showarticletitle{TutorialLens: Authoring Interactive Augmented Reality Tutorials Through Narration and Demonstration}. In \bibinfo{booktitle}{\emph{Proceedings of the 2021 ACM Symposium on Spatial User Interaction}} (Virtual Event, USA) \emph{(\bibinfo{series}{SUI '21})}. \bibinfo{publisher}{Association for Computing Machinery}, \bibinfo{address}{New York, NY, USA}, Article \bibinfo{articleno}{16}, \bibinfo{numpages}{11}~pages.
\newblock
\showISBNx{9781450390910}
\href{https://doi.org/10.1145/3485279.3485289}{doi:\nolinkurl{10.1145/3485279.3485289}}


\bibitem[Langlotz et~al\mbox{.}(2013)]%
        {mobileaudioanno13}
\bibfield{author}{\bibinfo{person}{Tobias Langlotz}, \bibinfo{person}{Holger Regenbrecht}, \bibinfo{person}{Stefanie Zollmann}, {and} \bibinfo{person}{Dieter Schmalstieg}.} \bibinfo{year}{2013}\natexlab{}.
\newblock \showarticletitle{Audio stickies: visually-guided spatial audio annotations on a mobile augmented reality platform}. In \bibinfo{booktitle}{\emph{Proceedings of the 25th Australian Computer-Human Interaction Conference: Augmentation, Application, Innovation, Collaboration}} (Adelaide, Australia) \emph{(\bibinfo{series}{OzCHI '13})}. \bibinfo{publisher}{Association for Computing Machinery}, \bibinfo{address}{New York, NY, USA}, \bibinfo{pages}{545–554}.
\newblock
\showISBNx{9781450325257}
\href{https://doi.org/10.1145/2541016.2541022}{doi:\nolinkurl{10.1145/2541016.2541022}}


\bibitem[Lee et~al\mbox{.}(2025)]%
        {imaginatear25}
\bibfield{author}{\bibinfo{person}{Jaewook Lee}, \bibinfo{person}{Filippo Aleotti}, \bibinfo{person}{Diego Mazala}, \bibinfo{person}{Guillermo Garcia-Hernando}, \bibinfo{person}{Sara Vicente}, \bibinfo{person}{Oliver~James Johnston}, \bibinfo{person}{Isabel Kraus-Liang}, \bibinfo{person}{Jakub Powierza}, \bibinfo{person}{Donghoon Shin}, \bibinfo{person}{Jon~E. Froehlich}, \bibinfo{person}{Gabriel Brostow}, {and} \bibinfo{person}{Jessica Van~Brummelen}.} \bibinfo{year}{2025}\natexlab{}.
\newblock \showarticletitle{ImaginateAR: AI-Assisted In-Situ Authoring in Augmented Reality}. In \bibinfo{booktitle}{\emph{Proceedings of the 38th Annual ACM Symposium on User Interface Software and Technology}} \emph{(\bibinfo{series}{UIST '25})}. \bibinfo{publisher}{Association for Computing Machinery}, \bibinfo{address}{New York, NY, USA}, Article \bibinfo{articleno}{52}, \bibinfo{numpages}{21}~pages.
\newblock
\showISBNx{9798400720376}
\href{https://doi.org/10.1145/3746059.3747635}{doi:\nolinkurl{10.1145/3746059.3747635}}


\bibitem[Li et~al\mbox{.}(2025)]%
        {satori25}
\bibfield{author}{\bibinfo{person}{Chenyi Li}, \bibinfo{person}{Guande Wu}, \bibinfo{person}{Gromit Yeuk-Yin Chan}, \bibinfo{person}{Dishita~Gdi Turakhia}, \bibinfo{person}{Sonia Castelo~Quispe}, \bibinfo{person}{Dong Li}, \bibinfo{person}{Leslie Welch}, \bibinfo{person}{Claudio Silva}, {and} \bibinfo{person}{Jing Qian}.} \bibinfo{year}{2025}\natexlab{}.
\newblock \showarticletitle{Satori: Towards Proactive AR Assistant with Belief-Desire-Intention User Modeling}. In \bibinfo{booktitle}{\emph{Proceedings of the 2025 CHI Conference on Human Factors in Computing Systems}} \emph{(\bibinfo{series}{CHI '25})}. \bibinfo{publisher}{Association for Computing Machinery}, \bibinfo{address}{New York, NY, USA}, Article \bibinfo{articleno}{1229}, \bibinfo{numpages}{24}~pages.
\newblock
\showISBNx{9798400713941}
\href{https://doi.org/10.1145/3706598.3714188}{doi:\nolinkurl{10.1145/3706598.3714188}}


\bibitem[Liu et~al\mbox{.}(2023)]%
        {instruar23}
\bibfield{author}{\bibinfo{person}{Ziyi Liu}, \bibinfo{person}{Zhengzhe Zhu}, \bibinfo{person}{Enze Jiang}, \bibinfo{person}{Feichi Huang}, \bibinfo{person}{Ana~M Villanueva}, \bibinfo{person}{Xun Qian}, \bibinfo{person}{Tianyi Wang}, {and} \bibinfo{person}{Karthik Ramani}.} \bibinfo{year}{2023}\natexlab{}.
\newblock \showarticletitle{InstruMentAR: Auto-Generation of Augmented Reality Tutorials for Operating Digital Instruments Through Recording Embodied Demonstration}. In \bibinfo{booktitle}{\emph{Proceedings of the 2023 CHI Conference on Human Factors in Computing Systems}} (Hamburg, Germany) \emph{(\bibinfo{series}{CHI '23})}. \bibinfo{publisher}{Association for Computing Machinery}, \bibinfo{address}{New York, NY, USA}, Article \bibinfo{articleno}{32}.
\newblock
\showISBNx{9781450394215}
\href{https://doi.org/10.1145/3544548.3581442}{doi:\nolinkurl{10.1145/3544548.3581442}}


\bibitem[Mohr et~al\mbox{.}(2015)]%
        {retarget15}
\bibfield{author}{\bibinfo{person}{Peter Mohr}, \bibinfo{person}{Bernhard Kerbl}, \bibinfo{person}{Michael Donoser}, \bibinfo{person}{Dieter Schmalstieg}, {and} \bibinfo{person}{Denis Kalkofen}.} \bibinfo{year}{2015}\natexlab{}.
\newblock \showarticletitle{Retargeting Technical Documentation to Augmented Reality}. In \bibinfo{booktitle}{\emph{Proceedings of the 33rd Annual ACM Conference on Human Factors in Computing Systems}} (Seoul, Republic of Korea) \emph{(\bibinfo{series}{CHI '15})}. \bibinfo{publisher}{Association for Computing Machinery}, \bibinfo{address}{New York, NY, USA}, \bibinfo{pages}{3337–3346}.
\newblock
\showISBNx{9781450331456}
\href{https://doi.org/10.1145/2702123.2702490}{doi:\nolinkurl{10.1145/2702123.2702490}}


\bibitem[Mohr et~al\mbox{.}(2017)]%
        {videoar17}
\bibfield{author}{\bibinfo{person}{Peter Mohr}, \bibinfo{person}{David Mandl}, \bibinfo{person}{Markus Tatzgern}, \bibinfo{person}{Eduardo Veas}, \bibinfo{person}{Dieter Schmalstieg}, {and} \bibinfo{person}{Denis Kalkofen}.} \bibinfo{year}{2017}\natexlab{}.
\newblock \showarticletitle{Retargeting Video Tutorials Showing Tools With Surface Contact to Augmented Reality}. In \bibinfo{booktitle}{\emph{Proceedings of the 2017 CHI Conference on Human Factors in Computing Systems}} (Denver, Colorado, USA) \emph{(\bibinfo{series}{CHI '17})}. \bibinfo{publisher}{Association for Computing Machinery}, \bibinfo{address}{New York, NY, USA}, \bibinfo{pages}{6547–6558}.
\newblock
\showISBNx{9781450346559}
\href{https://doi.org/10.1145/3025453.3025688}{doi:\nolinkurl{10.1145/3025453.3025688}}


\bibitem[Nguyen et~al\mbox{.}(2018)]%
        {asmim18}
\bibfield{author}{\bibinfo{person}{Tam~V. Nguyen}, \bibinfo{person}{Bilal Mirza}, \bibinfo{person}{Dorothy Tan}, {and} \bibinfo{person}{Jose Sepulveda}.} \bibinfo{year}{2018}\natexlab{}.
\newblock \showarticletitle{ASMIM: Augmented Reality Authoring System for Mobile Interactive Manuals}. In \bibinfo{booktitle}{\emph{Proceedings of the 12th International Conference on Ubiquitous Information Management and Communication}} (Langkawi, Malaysia) \emph{(\bibinfo{series}{IMCOM '18})}. \bibinfo{publisher}{Association for Computing Machinery}, \bibinfo{address}{New York, NY, USA}, Article \bibinfo{articleno}{3}, \bibinfo{numpages}{6}~pages.
\newblock
\showISBNx{9781450363853}
\href{https://doi.org/10.1145/3164541.3164592}{doi:\nolinkurl{10.1145/3164541.3164592}}


\bibitem[Poretski and Tang(2022)]%
        {gamelearn22}
\bibfield{author}{\bibinfo{person}{Lev Poretski} {and} \bibinfo{person}{Anthony Tang}.} \bibinfo{year}{2022}\natexlab{}.
\newblock \showarticletitle{Press A to Jump: Design Strategies for Video Game Learnability}. In \bibinfo{booktitle}{\emph{Proceedings of the 2022 CHI Conference on Human Factors in Computing Systems}} (New Orleans, LA, USA) \emph{(\bibinfo{series}{CHI '22})}. \bibinfo{publisher}{Association for Computing Machinery}, \bibinfo{address}{New York, NY, USA}, Article \bibinfo{articleno}{155}, \bibinfo{numpages}{26}~pages.
\newblock
\showISBNx{9781450391573}
\href{https://doi.org/10.1145/3491102.3517685}{doi:\nolinkurl{10.1145/3491102.3517685}}


\bibitem[Seo et~al\mbox{.}(2024)]%
        {gradual24}
\bibfield{author}{\bibinfo{person}{HyunA Seo}, \bibinfo{person}{Juheon Yi}, \bibinfo{person}{Rajesh Balan}, {and} \bibinfo{person}{Youngki Lee}.} \bibinfo{year}{2024}\natexlab{}.
\newblock \showarticletitle{GradualReality: Enhancing Physical Object Interaction in Virtual Reality via Interaction State-Aware Blending}. In \bibinfo{booktitle}{\emph{Proceedings of the 37th Annual ACM Symposium on User Interface Software and Technology}} (Pittsburgh, PA, USA) \emph{(\bibinfo{series}{UIST '24})}. \bibinfo{publisher}{Association for Computing Machinery}, \bibinfo{address}{New York, NY, USA}, Article \bibinfo{articleno}{82}, \bibinfo{numpages}{14}~pages.
\newblock
\showISBNx{9798400706288}
\href{https://doi.org/10.1145/3654777.3676463}{doi:\nolinkurl{10.1145/3654777.3676463}}


\bibitem[Shi et~al\mbox{.}(2025)]%
        {caring-ai25}
\bibfield{author}{\bibinfo{person}{Jingyu Shi}, \bibinfo{person}{Rahul Jain}, \bibinfo{person}{Seunggeun Chi}, \bibinfo{person}{Hyungjun Doh}, \bibinfo{person}{Hyung-gun Chi}, \bibinfo{person}{Alexander~J. Quinn}, {and} \bibinfo{person}{Karthik Ramani}.} \bibinfo{year}{2025}\natexlab{}.
\newblock \showarticletitle{CARING-AI: Towards Authoring Context-aware Augmented Reality INstruction through Generative Artificial Intelligence}. In \bibinfo{booktitle}{\emph{Proceedings of the 2025 CHI Conference on Human Factors in Computing Systems}} \emph{(\bibinfo{series}{CHI '25})}. \bibinfo{publisher}{Association for Computing Machinery}, \bibinfo{address}{New York, NY, USA}, Article \bibinfo{articleno}{29}, \bibinfo{numpages}{23}~pages.
\newblock
\showISBNx{9798400713941}
\href{https://doi.org/10.1145/3706598.3713348}{doi:\nolinkurl{10.1145/3706598.3713348}}


\bibitem[Srinidhi et~al\mbox{.}(2024)]%
        {xair24}
\bibfield{author}{\bibinfo{person}{Sruti Srinidhi}, \bibinfo{person}{Edward Lu}, {and} \bibinfo{person}{Anthony Rowe}.} \bibinfo{year}{2024}\natexlab{}.
\newblock \showarticletitle{XaiR: An XR Platform that Integrates Large Language Models with the Physical World}. In \bibinfo{booktitle}{\emph{2024 IEEE International Symposium on Mixed and Augmented Reality (ISMAR)}}. \bibinfo{pages}{759--767}.
\newblock
\href{https://doi.org/10.1109/ISMAR62088.2024.00091}{doi:\nolinkurl{10.1109/ISMAR62088.2024.00091}}


\bibitem[Stover and Bowman(2024)]%
        {TARGAR24}
\bibfield{author}{\bibinfo{person}{Daniel Stover} {and} \bibinfo{person}{Doug Bowman}.} \bibinfo{year}{2024}\natexlab{}.
\newblock \showarticletitle{TAGGAR: General-Purpose Task Guidance from Natural Language in Augmented Reality using Vision-Language Models}. In \bibinfo{booktitle}{\emph{Proceedings of the 2024 ACM Symposium on Spatial User Interaction}} (Trier, Germany) \emph{(\bibinfo{series}{SUI '24})}. \bibinfo{publisher}{Association for Computing Machinery}, \bibinfo{address}{New York, NY, USA}, Article \bibinfo{articleno}{12}, \bibinfo{numpages}{12}~pages.
\newblock
\showISBNx{9798400710889}
\href{https://doi.org/10.1145/3677386.3682095}{doi:\nolinkurl{10.1145/3677386.3682095}}


\bibitem[Tang et~al\mbox{.}(2025)]%
        {llmarreview25}
\bibfield{author}{\bibinfo{person}{Yiliu Tang}, \bibinfo{person}{Jason Situ}, \bibinfo{person}{Andrea~Yaoyun Cui}, \bibinfo{person}{Mengke Wu}, {and} \bibinfo{person}{Yun Huang}.} \bibinfo{year}{2025}\natexlab{}.
\newblock \showarticletitle{LLM Integration in Extended Reality: A Comprehensive Review of Current Trends, Challenges, and Future Perspectives}. In \bibinfo{booktitle}{\emph{Proceedings of the 2025 CHI Conference on Human Factors in Computing Systems}} \emph{(\bibinfo{series}{CHI '25})}. \bibinfo{publisher}{Association for Computing Machinery}, \bibinfo{address}{New York, NY, USA}, Article \bibinfo{articleno}{1054}, \bibinfo{numpages}{24}~pages.
\newblock
\showISBNx{9798400713941}
\href{https://doi.org/10.1145/3706598.3714224}{doi:\nolinkurl{10.1145/3706598.3714224}}


\bibitem[Thoravi~Kumaravel et~al\mbox{.}(2019)]%
        {loki19}
\bibfield{author}{\bibinfo{person}{Balasaravanan Thoravi~Kumaravel}, \bibinfo{person}{Fraser Anderson}, \bibinfo{person}{George Fitzmaurice}, \bibinfo{person}{Bjoern Hartmann}, {and} \bibinfo{person}{Tovi Grossman}.} \bibinfo{year}{2019}\natexlab{}.
\newblock \showarticletitle{Loki: Facilitating Remote Instruction of Physical Tasks Using Bi-Directional Mixed-Reality Telepresence}. In \bibinfo{booktitle}{\emph{Proceedings of the 32nd Annual ACM Symposium on User Interface Software and Technology}} (New Orleans, LA, USA) \emph{(\bibinfo{series}{UIST '19})}. \bibinfo{publisher}{Association for Computing Machinery}, \bibinfo{address}{New York, NY, USA}, \bibinfo{pages}{161–174}.
\newblock
\showISBNx{9781450368162}
\href{https://doi.org/10.1145/3332165.3347872}{doi:\nolinkurl{10.1145/3332165.3347872}}


\bibitem[Tran et~al\mbox{.}(2025)]%
        {articulate25}
\bibfield{author}{\bibinfo{person}{Nhan~(Nathan) Tran}, \bibinfo{person}{Ethan Yang}, {and} \bibinfo{person}{Abe Davis}.} \bibinfo{year}{2025}\natexlab{}.
\newblock \showarticletitle{ARticulate: Interactive Visual Guidance for Demonstrated Rotational Degrees of Freedom in Mobile AR}. In \bibinfo{booktitle}{\emph{Proceedings of the 2025 CHI Conference on Human Factors in Computing Systems}} \emph{(\bibinfo{series}{CHI '25})}. \bibinfo{publisher}{Association for Computing Machinery}, \bibinfo{address}{New York, NY, USA}, Article \bibinfo{articleno}{28}, \bibinfo{numpages}{8}~pages.
\newblock
\showISBNx{9798400713941}
\href{https://doi.org/10.1145/3706598.3713179}{doi:\nolinkurl{10.1145/3706598.3713179}}


\bibitem[Woodward and Ruiz(2023)]%
        {arsituaware23}
\bibfield{author}{\bibinfo{person}{Julia Woodward} {and} \bibinfo{person}{Jaime Ruiz}.} \bibinfo{year}{2023}\natexlab{}.
\newblock \showarticletitle{Analytic Review of Using Augmented Reality for Situational Awareness}.
\newblock \bibinfo{journal}{\emph{IEEE Transactions on Visualization and Computer Graphics}} \bibinfo{volume}{29}, \bibinfo{number}{4} (\bibinfo{year}{2023}), \bibinfo{pages}{2166--2183}.
\newblock
\href{https://doi.org/10.1109/TVCG.2022.3141585}{doi:\nolinkurl{10.1109/TVCG.2022.3141585}}


\bibitem[Yamaguchi et~al\mbox{.}(2020)]%
        {videoar20}
\bibfield{author}{\bibinfo{person}{Masahiro Yamaguchi}, \bibinfo{person}{Shohei Mori}, \bibinfo{person}{Peter Mohr}, \bibinfo{person}{Markus Tatzgern}, \bibinfo{person}{Ana Stanescu}, \bibinfo{person}{Hideo Saito}, {and} \bibinfo{person}{Denis Kalkofen}.} \bibinfo{year}{2020}\natexlab{}.
\newblock \showarticletitle{Video-Annotated Augmented Reality Assembly Tutorials}. In \bibinfo{booktitle}{\emph{Proceedings of the 33rd Annual ACM Symposium on User Interface Software and Technology}} (Virtual Event, USA) \emph{(\bibinfo{series}{UIST '20})}. \bibinfo{publisher}{Association for Computing Machinery}, \bibinfo{address}{New York, NY, USA}, \bibinfo{pages}{1010–1022}.
\newblock
\showISBNx{9781450375146}
\href{https://doi.org/10.1145/3379337.3415819}{doi:\nolinkurl{10.1145/3379337.3415819}}


\bibitem[Yousri et~al\mbox{.}(2024)]%
        {yousri2024illusionx}
\bibfield{author}{\bibinfo{person}{Ramez Yousri}, \bibinfo{person}{Zeyad Essam}, \bibinfo{person}{Yehia Kareem}, \bibinfo{person}{Youstina Sherief}, \bibinfo{person}{Sherry Gamil}, {and} \bibinfo{person}{Soha Safwat}.} \bibinfo{year}{2024}\natexlab{}.
\newblock \showarticletitle{IllusionX: An LLM-powered mixed reality personal companion}.
\newblock \bibinfo{journal}{\emph{arXiv preprint arXiv:2402.07924}} (\bibinfo{year}{2024}).
\newblock


\bibitem[{yt-dlp contributors}(2026)]%
        {ytdlp}
\bibfield{author}{\bibinfo{person}{{yt-dlp contributors}}.} \bibinfo{year}{2026}\natexlab{}.
\newblock \bibinfo{title}{yt-dlp}.
\newblock \bibinfo{howpublished}{\url{https://github.com/yt-dlp/yt-dlp}}.
\newblock
\newblock
\shownote{Software repository. Accessed March 30, 2026}.


\bibitem[Yu et~al\mbox{.}(2024)]%
        {feedforward24}
\bibfield{author}{\bibinfo{person}{Xingyao Yu}, \bibinfo{person}{Benjamin Lee}, {and} \bibinfo{person}{Michael Sedlmair}.} \bibinfo{year}{2024}\natexlab{}.
\newblock \showarticletitle{Design Space of Visual Feedforward And Corrective Feedback in XR-Based Motion Guidance Systems}. In \bibinfo{booktitle}{\emph{Proceedings of the 2024 CHI Conference on Human Factors in Computing Systems}} (Honolulu, HI, USA) \emph{(\bibinfo{series}{CHI '24})}. \bibinfo{publisher}{Association for Computing Machinery}, \bibinfo{address}{New York, NY, USA}, Article \bibinfo{articleno}{723}, \bibinfo{numpages}{15}~pages.
\newblock
\showISBNx{9798400703300}
\href{https://doi.org/10.1145/3613904.3642143}{doi:\nolinkurl{10.1145/3613904.3642143}}


\bibitem[Zhao et~al\mbox{.}(2025)]%
        {guided25}
\bibfield{author}{\bibinfo{person}{Ada~Yi Zhao}, \bibinfo{person}{Aditya Gunturu}, \bibinfo{person}{Ellen Yi-Luen Do}, {and} \bibinfo{person}{Ryo Suzuki}.} \bibinfo{year}{2025}\natexlab{}.
\newblock \showarticletitle{Guided Reality: Generating Visually-Enriched AR Task Guidance with LLMs and Vision Models}. In \bibinfo{booktitle}{\emph{Proceedings of the 38th Annual ACM Symposium on User Interface Software and Technology}} \emph{(\bibinfo{series}{UIST '25})}. \bibinfo{publisher}{Association for Computing Machinery}, \bibinfo{address}{New York, NY, USA}, Article \bibinfo{articleno}{146}, \bibinfo{numpages}{15}~pages.
\newblock
\showISBNx{9798400720376}
\href{https://doi.org/10.1145/3746059.3747784}{doi:\nolinkurl{10.1145/3746059.3747784}}


\bibitem[Zhu et~al\mbox{.}(2025)]%
        {agentar25}
\bibfield{author}{\bibinfo{person}{Chenfei Zhu}, \bibinfo{person}{Shao-Kang Hsia}, \bibinfo{person}{Xiyun Hu}, \bibinfo{person}{Ziyi Liu}, \bibinfo{person}{Jingyu Shi}, {and} \bibinfo{person}{Karthik Ramani}.} \bibinfo{year}{2025}\natexlab{}.
\newblock \showarticletitle{agentAR: Creating Augmented Reality Applications with Tool-Augmented LLM-based Autonomous Agents}. In \bibinfo{booktitle}{\emph{Proceedings of the 38th Annual ACM Symposium on User Interface Software and Technology}} \emph{(\bibinfo{series}{UIST '25})}. \bibinfo{publisher}{Association for Computing Machinery}, \bibinfo{address}{New York, NY, USA}, Article \bibinfo{articleno}{54}, \bibinfo{numpages}{23}~pages.
\newblock
\showISBNx{9798400720376}
\href{https://doi.org/10.1145/3746059.3747676}{doi:\nolinkurl{10.1145/3746059.3747676}}


\end{thebibliography}

\appendix

\section{Prompt Strategy}
\label{app:prompt}
For the prompt strategy, we explicitly constrain the model to produce valid JSON so that the output can be directly parsed by the client and mapped to rendering actions. This structured format also enables a clear separation between symbolic planning and situated execution: the planner specifies what should be done and what kind of guidance should be shown, while the runtime verifier determines whether the current step is complete and whether local refinement is needed.

\subsection{Initial Prompt}
In the initial plan, each step contains an instruction, one or more visualization types, a target object or region to localize, and a verification rule that defines success.

\begin{promptbox}
**Output valid JSON only, do not include extra text or comments.**

goal: One sentence starting with a verb, clear and testable (e.g., "Boil a pot of water").

steps:
Break into 3–12 steps; each starts with a verb and uses consistent granularity.

plannerResponse.next:
Must be one of the steps, or a sub-step of a step (e.g., "Heat the pot / set to medium heat").

viz:
Each step requires TWO types of visualization:

1. objectViz: How to visualize the object to operate on
   - "Outline": Green bounding box to identify object location (use when object position needs highlighting)
   - "ShapePreview": Shape/area preview image (use when object shape or area needs to be shown, especially with SAM3 segmentation)

2. actionViz: How to visualize the hand/target object action or movement
   - "Arrow": For movement/rotation (translation or rotation). Requires waypoints with start and end positions.
   - "Gesture": For hand gestures (e.g., pinch, poke, grip). System will search for matching image in Resources/Image.
   - "Tool": For tool actions (e.g., mouseclick). System will search for matching image in Resources/Image.

SPECIAL RULE: If both objectViz="ShapePreview" AND actionViz="Arrow", the system will animate the ShapePreview image instead of showing an arrow.

waypoints:
At least one "target" waypoint for the object to operate on.

If actionViz="Arrow" with translation, include both "target" (start) and "endtarget" (end position).

If actionViz="Arrow" with rotation, include "target" with rotation axis and direction information.

The object name should be precise, and different types of names can be explored. Adjectives may be added to further describe the object (e.g., white knob, black pot). 

CRITICAL: 
   
   - If tracking an item in the game world, name it directly (e.g., "flint stone on ground").
   
   - If tracking text or UI, use the word "text" or "button" (e.g., "Pick up flint text", "Crafting button", "Axe icon in Leftside UI").
   
   - If tracking keyboard, use the description like "W on keyboard".
   
   - NEVER use ambiguous words like "prompt", "area" as the objectName.
\end{promptbox}
\begin{jsonbox}
json format: 
{
  "goal": "",
  "steps": [],
  "plannerResponse": {
    "next": "",
    "check": "",
    "success": false,
    "viz": {
      "objectViz": "Outline|ShapePreview",
      "actionViz": "Arrow|Gesture|Tool|null",
      "actionType": ["translation"] | ["rotation"] | ["pinch"] | ["mouseclick"] | ...,
      "needsTranslation": true|false,
      "needsRotation": true|false,
      "waypoints": [
        {
          "type": "target",
          "objectName": ""
        },
        {
          "type": "endtarget",
          "objectName": ""
        }
      ]
    }
  }
}

Example:
{
  "goal": "Turn the gas knob to medium heat",
  "steps": [
    "Locate the gas knob",
    "Turn the gas knob clockwise to medium position"
  ],
  "plannerResponse": {
    "next": "Turn the gas knob clockwise to medium position",
    "check": "",
    "success": false,
    "viz": {
      "objectViz": "Outline",
      "actionViz": "Arrow",
      "actionType": ["rotation"],
      "needsRotation": true,
      "needsTranslation": false,
      "waypoints": [
        {
          "type": "target",
          "objectName": "silver gas knob"
        }
      ]
    }
  }
}
\end{jsonbox}

\subsection{During-task Prompt}
During execution, the verification output augments this structure with completion status, error feedback, and optional substeps.

\begin{promptbox}
You are a task tutor. You will be given an existing plannerResponse as a template. Your task is to:

1. FILL IN these three fields based on the current photo:

   - next: The specific sub-goal for the next step (if current step is not successfully reached).
   
   - success: Please check the image and confirm if current step is reached, only answer true or false.
   
   - check: If you are not sure of the result of success, tell me what you need to further check. If you are sure, leave it empty or skip it.

2. VERIFY the field from the existing plannerResponse:

   - waypoints: Verify that the waypoint objectNames and types are still relevant and suitable for current visualization. Keep the existing waypoint structure unless it's completely inappropriate.

IMPORTANT: Use the existing plannerResponse as your base template. Ignore any unnecessary details when judging the status. e.g. only focus on paper shape when doing origami.

\end{promptbox}

\subsection{Rotation Prompt}

\begin{promptbox}
The pivot point will rotate around a pivot point ({objectName}) in the image to match the last reference image. Please identify the position of pivot point. Refer to the first image to set up the spatial axis, with X pointing rightward in the photo,  Y pointing physically upward, and Z pointing toward you. Identify the rotational axis as X, Y, or Z. Always look from the positive side and determine the rotation direction as Positive (clockwise) or Negative (CounterClockwise). Each value must be int between 0–1000 normalized. Output using this JSON format:     
\end{promptbox}
\begin{jsonbox}
{{name: "", pos: [x_min, y_min, x_max, y_max], rotation: [axis, direction]}}
\end{jsonbox}
\subsection{Transformation Prompt}

\begin{promptbox}
    Please identify where {objectName} will end up after the in-step guidance. Each value must be int between 0-1000 normalized. Output each position item using this JSON format: 
\end{promptbox}
\begin{jsonbox}
{type: "starttarget | endtarget | object", name: "exact waypoint name", pos: [x_min, y_min, x_max, y_max]}.
\end{jsonbox}

\end{document}